\documentstyle[epsfig,12pt]{article}
\newcommand{\ds}{\displaystyle}
\newcommand{\beq}{\begin{equation}}
\newcommand{\eeq}{\end{equation}}
\newcommand{\bea}{\begin{eqnarray}}
\newcommand{\eea}{\end{eqnarray}}
\newcommand{\Lam}{\Lambda_{\overline{MS}}}
\newcommand{\eps}{\epsilon}
\newcommand{\MS}{\overline{MS}}
\newcommand{\qq}{\langle \bar q q \rangle}

\newcommand{\nn}{\nonumber}

\begin{document}
\begin{titlepage} 
\title{\vspace{-15mm}
       {\normalsize \hfill
       \begin{tabbing}
       \`\begin{tabular}{l}
  CERN--TH/96--243 \\
         hep-ph/9609265\\
        \end{tabular}
       \end{tabbing} }
       \vspace{8mm}
Variational Quark Mass Expansion and the \\ Order Parameters
of Chiral Symmetry Breaking  } 
\author{Jean-Lo\"{\i}c Kneur\thanks{On leave
from Laboratoire de Physique Math\'ematique et Th\'eorique,
U.R.A. 768 du C.N.R.S.,  F34095 Montpellier Cedex 5, France.}}
\date{}
\maketitle
\vspace{-1cm}
\begin{center}
{\em CERN, Theoretical Physics Division \\
CH-1211 Geneva 23 Switzerland }
\end{center}
\begin{abstract}
\setlength{\baselineskip}{20pt}
We investigate in some detail
a ``variational mass" expansion approach,
generalized from a similar construction developed in 
the Gross-Neveu model, to evaluate
the basic order parameters
of the dynamical breaking of the $SU(2)_L \times SU(2)_R$
and $SU(3)_L \times SU(3)_R$ chiral
symmetries in QCD. The method starts with a reorganization of
the ordinary perturbation theory with the addition of
an {\em arbitrary} quark mass $m$.
The new perturbative series
can be summed to all orders thanks to
renormalization group properties, with specific
boundary conditions,
and advocated analytic continuation in $m$ properties.
In the approximation
where the explicit breakdown of the chiral
symmetries due to small current quark masses is neglected,
we derive 
ansatzes
for the
dynamical contribution to the
``constituent" masses $M_q$ of the $u,d,s$ quarks;
the pion decay
constant $F_\pi$; and the quark condensate $\qq$
in terms of the
basic QCD scale
$\Lam $. Those ansatzes are then
optimized, in a sense to be specified, and also explicit symmetry breaking
mass terms can be consistently introduced in the framework.
The values of $F_\pi$ and $M_q$
obtained are roughly in agreement
with what is expected from other non-perturbative
methods. In contrast we obtain quite a small value of
$\vert \qq \vert $ within our
approach. The
possible interpretation
of the latter results is briefly discussed.
\end{abstract}
{\footnotesize PACS No 12.38Aw, 12.38Lg, 11.30.Rd, 11.15.Bt}\\ 
CERN--TH/96--243 \\
August 1996
\end{titlepage}
\vfill

\setlength{\baselineskip}{15pt}
\def\baselinestretch{1.}
\section{Introduction}
\setcounter{footnote}{0}
Although QCD is certainly established as the correct theoretical framework
underlying the strong interaction physics,
a still challenging issue is to derive the 
low-energy properties of the strongly interacting spectrum from 
QCD ``first principle",
due to our limited present skill with non-perturbative technics.
At very low energy, where the (ordinary) 
perturbation theory cannot be applied, 
Chiral Perturbation theory (ChPT)~\cite{GaLeut} 
or extended Nambu--Jona-Lasinio models (ENJL)~\cite{NJL,ENJL}
give a consistent framework 
in terms of basic parameters that have to be fixed from the data.
Yet the link between those effective parameters and the basic
QCD ones, the gauge coupling and the quark masses, 
remains largely unsolved.
To have, typically, a determination of the pion decay constant 
$F_\pi$ and other 
similar low energy quantities 
directly in terms of the basic QCD coupling constant
$\alpha_S$, is clearly a desirable task. 
Lattice simulations
are perhaps at present
the only systematic approach to such questions, and 
indeed
provide, among many other things,
a determination of $\alpha_S$~\cite{lattalphas} and
also evidence for dynamical chiral symmetry breakdown 
(CSB)~\cite{lattcsb}.
Yet a 
 fully consistent treatment of the chiral symmetry on the lattice
is still missing, and there are also inherent difficulties of dealing
with truly dynamical ``unquenched" quarks, light meson masses,
and related problems. \\

In the present paper,
we investigate
a new method to implement CSB {\it directly} from the basic
QCD degrees of freedom, the quarks and gluons.
More precisely we shall explore how far
the basic QCD Lagrangian can provide, in a self-consistent
way, at least the basic
CSB parameters, namely the formation of
dynamical quark
masses, quark condensates, and the pion decay constant,
in the limit of vanishing Lagrangian (current) quark masses. Such a
qualitative picture
of CSB can be made more quantitative
by applying
a new ``variational mass" approach,
recently developed within the framework of the
anharmonic oscillator~\cite{bgn}, and 
the Gross-Neveu (GN) model~\cite{gn1,gn2}. \\ 

Before developing our construction, let us remind that 
there are in fact two main phenomena which are
believed to emerge from non-perturbative dynamics
(and which are not fully understood from first principle),
namely the Confinement
and CSB respectively.
Even if those are certainly intimately related in the full
QCD dynamics,
it is legitimate to consider,
at least in a first approximation, those two issues separately.
QCD is a very rich
theory involving many different non-perturbative aspects, therefore any
new method or model can most likely deal with a selected or simplified
aspect of the full dynamics. After all, {\it as far}
as the gross
features of the chiral dynamics are concerned (that is,
the existence of
light pions and consequences in the form of 
low energy theorems~\cite{LET}),
a picture without confinement but with CSB would
not look very much different from the true QCD one.
On the other hand, there is a large evidence that the confinement forces
do play a more major role in the formation of heavier hadrons~\cite{shuryak}.
This relatively neat separation of the relevant scales
is indeed one of the basic assumptions
underlying several successful approaches to
chiral dynamics, like the above-mentioned ChPT,
ENJL models etc,
which generally do not include the confinement properties in
their framework.
Somewhat closer to the present investigation, 
similar assumptions are often made in many dynamical CSB
models, dealing with various levels of approximations in the
treatment of the Schwinger-Dyson
equations for the Green functions relevant to chiral dynamics~\footnote{
See e.g. ref.~\cite{nagoya}
and the textbook by V. A. Miransky~\cite{Miransky} for 
reviews and original references.}.
In contrast, there exists
a quite radically different attitude towards CSB in QCD, advocating
that the responsible mechanism is most probably the non-perturbative
effects due to the {\em instanton} vacuum~\cite{CaDaGro},
or even more directly related to confinement~\cite{Cornwall}.
However,
even if the instanton picture of CSB is on general grounds well motivated,
and many fruitful ideas have been developed in that
context~\footnote{For a review on such issues and an extended collection
of original references, we refer to the textbook by
E. V. Shuryak~\cite{shuryak}.},
as far as we are aware there is at present no
sufficiently rigorous or compelling evidence for it.
In any event, it is certainly of interest to
investigate quantitatively
the ``non-instantonic" contribution to CSB, and we hope
that our method is a more consistent step in that direction. \\

Although the basic idea was already developed in refs~\cite{gn1,gn2} 
(see also \cite{qcd1} for a first acquaintance in QCD),
we shall reformulate here the construction in some details for
self-containedness reasons, 
and to take into account some crucial differences  
between QCD and the GN model. In addition we introduce 
explicit symmetry breaking mass terms consistently in the framework, 
which as we shall see play an important role in the determination
of $\qq$. \\
The method thus starts by considering an {\it arbitrary}
quark mass term $m$ added to the massless QCD Lagrangian,
whose dependence is calculated in perturbation theory with an
expansion parameter $x$,
interpolating between the massive free theory, at $x=0$,
and the relevant interacting theory, at $x=1$. This starting point is  
similar to
the one developed a long time ago and implemented 
in various different forms in refs.\cite{pms}--\cite{SOLO}. 
There, it was advocated 
that the convergence of conventional perturbation
theory may be improved by such a variational procedure in which
the separation of the
action into ``free" and ``interaction" parts
is made to depend on some set of auxiliary parameters.
The results obtained  by expanding to finite order
 in this redefined perturbation
series are optimal in regions  of the  space of auxiliary
parameters where they are least sensitive to these parameters.
Indeed, such regions are those expected to
best approximate the exact answer, where there should
simply be {\em no} dependence on the auxiliary 
parameters~\footnote{For variants of the optimized 
variational-perturbation theory ideas,
not based on minimization, but leading in many field-theoretical
applications to similar
results, see also refs.~\cite{Grunberg,BLM}.}.
Moreover it recently appeared
strong evidence that this optimized perturbation
theory may lead to a rigorously convergent
series of approximations even in strong coupling cases.
In particular, the convergence of this variational-like procedure has been
rigorously established in the case of zero
and one dimensional field theories\cite{JONES}.

An essential novelty
however is that the construction in \cite{gn1,gn2} 
combines in a specific manner 
the renormalization group (RG) invariance  
of the theory 
together with analytic continuation in $m$  
properties: this, at least in a certain approximation to be motivated, 
allows to reach 
{\it infinite} order of perturbation in the 
parameter $x$, therefore presumably optimal,
provided it converges. This leads to a set of
non-perturbative
ansatzs for the relevant CSB quantities, 
as functions of the variational mass parameter $m$ and 
renormalization scheme (RS) dependence parameters as well, 
which can be 
studied for extrema and optimized. 
Quite essentially, our construction also provides a simple and
consistent treatment of the renormalization, reconciling 
the variational approach with the inherent infinities
of quantum field theory and the RG properties.
A comparison of the variational calculations
with the known exact results~\cite{FNW}
 for the mass gap in the $O(N)$ GN model~\cite{gn} 
is very satisfactory\cite{gn2}, in view of the fact that
for arbitrary $N$ the (only known at present) second order 
non-logarithmic perturbative terms have been 
used in the optimization. 
For QCD, one might however argue that the truly 
non-perturbative effects are lying in the infrared
domain, for which a complete non-perturbative treatment is clearly 
beyond the scope of the
present framework. 
We shall see in fact that our 
RG invariant ansatz
can be viewed as an 
(optimized) perturbation theory 
around a non-trivial fixed 
point
of the RG evolution solutions, 
which in that sense 
avoids having to treat the infrared problems inherent to
non-perturbative QCD~\footnote{Quite recently, an interesting
approach has been developed~\cite{Wet}
to determine the same CSB quantities,
by using the exact ``renormalization group flow" 
formulation~\cite{WilKog}.
Due to the central role played by RG
properties, our present approach may
be considered as having some similarities in aim with the latter one,
albeit being totally different in the technics used.}. 
The physical relevance of such a framework  
can be judged a posteriori, e.g. by comparing our estimates
with the results from other non-perturbative methods (like the
lattice simulations~\cite{lattalphas,lattcsb} or 
the spectral sum rules approach~\cite{sumrules} 
typically), but unfortunately
there is at present no rigorous 
way of defining the {\em intrinsic} error and
convergence properties of the method, at least in four dimension.\\

The paper is organized as follows: in section 2, we 
explain and motivate in details the construction and 
derive the main ansatz 
for the dynamical CSB contribution to the ``constituent"
quark mass, 
in the exact chiral limit. 
A dynamical quark
mass in QCD is not, strictly speaking, an order parameter, 
but it is interesting for at least two reasons:  
first it appears as a direct generalization of 
the rather intuitive concept of a (fermion) mass
gap originating from dynamical CSB, 
as illustrated in
many simpler models. Second, as we shall see, 
the concept and known properties of a gauge-invariant and
infrared finite {\em pole} mass
enters as a basic ingredient in the 
ansatz, and plays also some role  
in the determination of the genuine 
order parameters, $\qq$ 
and $F_\pi $. Concisely, our construction will be shown to 
amount to: \\
{\it i}) taking the pole mass, as expressed as a function of the
Lagrangian (arbitrary) 
running mass parameter $m(\mu)$ in some appropriate renormalisation
scheme (RS); \\
\noindent 
{\it ii}) constructing an ansatz as a (contour) integral over 
the analytically continued mass parameter $m(\mu)$,
which can be
shown to formally resum the reorganized $x$ series,  
with a specific contour  
and appropriate RS
choices avoiding some singularities; \\
\noindent 
{\it iii}) finally optimizing the result with respect to $m(\mu)$ 
and the RS arbitrariness. \\ 
In section 3 the construction is generalized to the
case of a composite operator of arbitrary mass dimension $n$, 
and applied 
to the determination of the 
pion decay constant, $F_\pi $. 
In section 4 we apply the construction to 
derive from similar
arguments an expression for $\qq$, 
introducing for this purpose explicit symmetry
breaking mass terms. 
Summary and conclusions are 
given 
in section 5.
We have also collected in
Appendix A the relevant material on 
renormalization group properties, which fixes our normalizations 
and some definitions, and in Appendix B 
the analytic continuation technics used in the derivation of 
the main ansatzs.
For completeness we also  
display in Appendix C
the 
perturbative expressions which are used as basic ingredients in the  
different ansatzs.
\newpage 
\section{Dynamical quark masses}
\setcounter{equation}{0}
\subsection{Preliminaries: QCD Lagrangian with variational mass}
Let us start with the bare QCD Lagrangian 
(for $n_f$ quark fields $q_i$ of identical
masses)
\bea
L_{QCD} = -{1\over{4}}{G^{\mu \nu}_a G^a_{\mu \nu}}
+{\sum^{n_f}_{i=1} \overline{q}_i (i \gamma_\mu D^\mu -m_0) q_i}
\label{LQCD}
\eea
where
\bea
G^{\mu \nu}_a \equiv \partial^\mu G^\nu_a -\partial^\nu G^\mu_a -g_0 f^{abc}
G^\mu_b G^\nu_c\; ; \nn \\ 
D^\mu q_i \equiv \partial^\mu q_i +i g_0 T_a G^\mu_a q_i\; ;
\eea
are respectively the gluon field strength tensor and the covariant derivative,
$a$ is the color index, $i$ is flavor index, 
$T_a$ the $SU(3)_c$ generator, and $m_0$, $g_0$
the bare mass and gauge coupling constant, respectively.
As is well known the QCD lagrangian (\ref{LQCD}) 
possesses for $m_0 = 0$
the additional
chiral symmetry $SU(n_f)_L \times SU(n_f)_R \times U(1)_L \times
U(1)_R$, at the classical level~\footnote{We do not consider in particular
the effects
due to the breakdown of the axial $U(1)$ symmetry associated with 
instanton
phenomena~\cite{tHooft}.}.
In what follows we only consider the $SU(n_f)_L \times SU(n_f)_R$
part of the chiral symmetry, and for
$n_f =2$ or $n_f = 3$ as physically relevant applications.\\

Following the treatment of 
the GN model~\cite{gn2}, 
let us add and subtract a (bare) mass term,
$m_0 {\sum^{n_f}_{i=1}} \overline{q}_i q_i$,
to be treated as an interaction term,
to the {\it massless} QCD
Lagrangian. 
This is most conveniently done by introducing a new 
perturbation parameter $x$, 
according to
\beq
L_{QCD} \to L^{free}_{QCD}(g_0 =0, m_0=0)
-m_0 \sum^{n_f}_{i=1} \overline{q}_i q_i
+ L^{int}_{QCD}(g^2_0 \to x g^2_0)
+x \; m_0 \sum^{n_f}_{i=1} \overline{q}_i q_i
\label{xdef}
\eeq
where $L^{int}_{QCD}$ designates the usual QCD interaction terms, i.e
those proportional to $g_0$ or $g^2_0$ in (\ref{LQCD}).
This is formally equivalent 
to substitute everywhere in Lagrangian (\ref{LQCD})
\beq m_0 \to m_0\; (1-x); ~~~~g^2_0 \to g^2_0\; x, 
\label{substitution}
\eeq
and therefore 
in any perturbative (bare) quantity as well, 
calculated in terms of $m_0$ and $g_0$~\footnote{
It is implicitly understood that the same coupling constant 
reparametrization, (\ref{substitution}), also affects 
e.g. the Fadeev-Popov terms necessary at 
the loop-level calculations, which we will not display explicitly. 
Note therefore that the substitution (\ref{substitution}) does not
spoil the gauge-invariance.}.  
Since the massless 
Lagrangian is recovered for $x \to 1$, $m_0$ is to be considered as 
an arbitrary  mass parameter 
after substitution (\ref{substitution}). One expects
to optimize physical quantities with
respect to $m_0$ at different, possibly arbitrary 
orders of the expansion parameter $x$,
eventually approaching a stable limit, i.e {\it flattest}
 with respect to $m_0$,
 at sufficiently high order in $x$. \\
One immediately encounters
however a number of obstacles: first, 
and quite essentially, before accessing to any physical quantity
of interest for such an optimization, the theory has to be
renormalized, and there is an unevitable mismatch
between the expansion in $x$, as
introduced above, and the ordinary perturbative expansion as dictated by the
necessary counterterms, which at first spoils the whole picture.
Moreover, for any physical quantity of interest in QCD 
one only knows at present the first few terms of the ordinary
perturbative expansion in $g^2$, and the computational effort
required to perform perturbative calculation in $x$ is a priori
similar. Finally, even if arbitrary higher order
perturbation terms were available, it is easy to convince oneself that, 
at any finite order in the $x$ expansion,
one always 
recovers a trivial result in the limit $ m \to 0$ (equivalently 
$x\to 1$), which
is the limit we are ultimately 
interested in to identify a 
mass gap (and more generally to identify spontaneous chiral symmetry 
breakdown, since an explicit Lagrangian mass term breaks the axial vector
Ward-Takahashi identities explicitly~\cite{Miransky}).\\ 

All these 
problems can be circumvented at once, by advocating  a certain
resummation ansatz~\cite{gn2},
exploiting the RG  
properties of the theory 
plus analytic continuation properties  
in the arbitrary mass parameter $m$, 
which defines a candidate mass gap $M(m)$~\footnote{
We will introduce several different definitions of a quark mass. 
To avoid
confusion, we designate by $m$ the 
Lagrangian ``current" mass (or any mass trivially related to the latter)
, while $M$ (with possibly different 
indices) is used for  
{\em dynamical} mass candidates at different stages.}. 
With respect to ordinary perturbation theory this RG invariant
ansatz is 
exact for 
the leading and next-to-leading 
dependence in $m$ only, 
but the crucial point
is that it
resums as well the (reorganized) perturbation series in $x$.
Moreover, taking the limit $x \to 1$ 
in this resummed expression no longer gives
a zero mass gap. The construction can be systematically
improved order by order (at least in principle, 
whereas in practice we obviously limit ourselves 
to the presently known highest perturbative orders for the relevant 
quantities). This is nevertheless hoped to be 
a sufficiently good starting
point for further numerical optimization studies with respect to the
(arbitrary) parameter $m$~\footnote{ 
Such a situation is somewhat similar 
to many other variational approaches,
where one often starts
with a trial estimate, which one expects to be reasonably close to
the exact result, before performing a
variational calculation.}. 
For simplicity, we shall first 
illustrate the method by restricting ourselves in the next section 
to the first order of RG dependence. Generalization to higher orders
will be discussed in the next sections. 
 
\subsection{First order renormalization group mass ansatz}
To be meaningful our formalism should be applied to
RG invariant 
{\it and} gauge-invariant quantities, in the QCD context. 
Therefore 
a reasonable starting point 
is to consider the {\it pole} quark mass,
which is gauge-invariant~\cite{Tar} to arbitrary perturbative
orders~\cite{Lav}, and infrared finite~\cite{Tar}. 
Using dimensional regularisation
(with space-time dimension $ D = 4 -\eps$), the one-loop
expression for the quark mass, using the on-shell 
condition $ \not \! q = m_0$,  
is easily obtained
as
\beq
m_0~ [1 +3 C_F g^2_0 {\Gamma(\eps/2) \over{(4 \pi)^{2-\eps/2}}}
(m_0)^{-\eps} {( 1 +{2 \over 3} \eps +O(\eps^2))}\;+
O(g^4_0) ]\;,
\label{mpert01}
\eeq
where $C_F \equiv (N^2_c-1)/(2 N_c)$ for
$SU(N_c)$ color 
(i.e. $C_F = 4/3$ for QCD),  and $\Gamma(x)$ the Euler Gamma-function.
For $\eps \to 0$, expression (\ref{mpert01})
leads to the 
(renormalized) pole mass expression
\beq
M^P_1 \simeq \; \bar m~ (1 +2{\bar\alpha_S \over \pi}[\frac{2}{3} 
-\ln (\frac{\bar m}{\bar\mu})] 
+O(\alpha^2_S)~),
\label{mpertren}
\eeq
as expressed in terms of the $\MS$ mass 
$\bar m \equiv m(\bar\mu)$ 
($\bar \mu \equiv \mu \sqrt{4 \pi} e^{-\gamma_E/2}$, $\gamma_E \simeq
0.577216..$, where   
$\mu $ is the arbitrary scale introduced
by dimensional regularisation), 
and coupling $\bar\alpha_S \equiv g^2(\bar\mu)/(4 \pi)$.
In (\ref{mpertren}) we used 
the $\MS$ mass counterterm $Z_m$, defined as 
$m_0 \equiv Z_m \bar m$ and given from (\ref{mpert01}) as: 
$
Z_m (\bar g^2)  = 1 -{\bar g^2 /(2 \pi^2 \eps)} +O(g^4)$. 

Our construction starts by first 
observing that RG properties gives us more information
on higher orders than what appears in (\ref{mpert01}), 
(\ref{mpertren}):
one-loop calculations actually provide the leading 
$1/\epsilon$ dependence of $Z_m$ and of the counterterm for 
the coupling 
$Z_g$ to {\it all} orders, as is well known:
\beq
Z_m \equiv m_0/m 
= (1 +{2b_0 \over \epsilon} \bar g^2)^{
-\frac{\gamma_0}{2b_0}},
\;\;\; Z_g \equiv (g^2_0/g^2) \mu^{-\epsilon}
= (1 +{2b_0 \over \epsilon} \bar g^2)^{-1}\;,
\label{zmdef}
\eeq
where $\gamma_0$ and $b_0$ are the first order RG coefficient for the
anomalous mass dimension and beta function respectively (see Appendix A
for the normalization used). 

Now expression (\ref{mpert01}) using (\ref{zmdef})
suggests the 
following form of a {\it bare} resummed mass expression
(restricted at the moment to the first RG order):
\beq
{\ds M_1  = {m_0 \over{[1 -b_0 \Gamma(\epsilon /2) (4\pi)^{\epsilon/2}
g^2_0 
(M_1)^{-\epsilon} ]^{\gamma_0 \over{2b_0}} }}
},
\label{Mbareansatz1}
\eeq
in close analogy with the mass-gap formula of the GN
model~\cite{gn2}. Eq.~(\ref{Mbareansatz1}) is obviously RG invariant
since expressed only in terms of $m_0$ and $g_0$.
Introducing renormalized quantities $\bar m, \bar g^2$ from 
(\ref{zmdef}), 
(\ref{Mbareansatz1}) is in addition 
finite to {\it all } orders, as easily checked,
thanks to the recursivity in $M_1$,
and reads explicitly 
\beq
M_1 = {\bar m \over{[1 +2b_0 \bar g^2 \ln({M_1 \over{\bar \mu}})
]^{\gamma_0 \over{2b_0}}}}\;.
\label{Mrenansatz1}
\eeq
Eq.~(\ref{Mrenansatz1}) resums the
leading log (LL) dependence in $\bar m$ of the usual pole mass
but, evidently,
does not give its correct next-to-leading
dependence in $\bar m$ (nor even any of the 
``purely perturbative" (non-logarithmic) finite corrections, 
such as the 2/3 factor appearing already at first order
in (\ref{mpert01})--(\ref{mpertren})~). 
Actually, the renormalized form (\ref{Mrenansatz1}) can alternatively  
be rigorously derived as a particular
boundary condition imposed on the usual RG running  mass.
 Indeed, 
 consider the general
solution for the RG evolution of the running mass,  
\beq
m(\mu^{'}) = m(\mu )\;\; {exp\{ -\int^{g(\mu^{'})}_{g(\mu )}
dg {\gamma_m(g) \over {\beta(g)}} \} }
\label{runmass}
\eeq
in terms of the effective coupling $g(\mu)$, whose RG evolution is given
as $\mu dg(\mu)/d\mu \equiv \beta(g)$.
Solving (\ref{runmass}) with the following 
``fixed point" boundary condition:
\beq
M \equiv m(M),
\label{RGBC}
\eeq
one easily obtains (to first RG order)
\beq
M_1 = {m(\mu) \over{[1 +2b_0 g^2(\mu) \ln({M_1 \over{\mu}})
]^{\gamma_0 \over{2b_0}}}}\;, 
\label{MRG1}
\eeq
which is nothing but eq.~(\ref{Mrenansatz1}) for $\mu = \bar \mu$.
In particular, the recursive form of (\ref{MRG1}) with respect to $M_1$ can
only be obtained thanks to the specific boundary condition in 
(\ref{RGBC}). 
Note that, although (\ref{Mrenansatz1}) can thus be obtained 
solely from the condition
(\ref{RGBC}) imposed on the ``current" mass $m(\mu)$, it 
has the usual properties
of a pole mass~\footnote{In particular, $M_1$ in (\ref{MRG1}) 
is {\em scale} invariant, in contrast with $m(\mu)$:
i.e. an arbitrary change in $\mu$ in the current mass $m(\mu)$ 
is compensated, to all orders, by the $\ln \mu$ dependence in the 
denominator of (\ref{MRG1}). (\ref{MRG1}) is also clearly gauge-invariant, 
as the pole mass should be.}.
This coincidence between the 
 pole mass $M$ and the current mass
$m(M)$, evaluated at the scale $\mu \equiv M$, is of course only an artefact 
of our crude approximation, neglecting at the moment the non-logarithmic
perturbative corrections. 
 Although the first order 
expression (\ref{Mrenansatz1}) having those properties  
could  be guessed directly, the advantage of the latter derivation
 is that it gives a sound basis and precise link with
the renormalization group behaviour.
Indeed, as will be derived in the next section, 
one can still obtain an exact solution of (\ref{runmass}) with 
the boundary condition (\ref{RGBC}) at the next order explicitly,
which is by construction entirely consistent with the 
next-to-leading RG dependence, to all orders. We shall also 
include later on 
the non-logarithmic perturbative corrections, necessary to
make contact with the genuine pole mass and to define 
a more realistic ``mass gap" ansatz.\\ 

Now, the most important
property of the simple-minded expression (\ref{Mrenansatz1}), which is also 
shared by the more realistic ansatzs to be later derived, 
is that it is 
{\em non-zero} in the chiral limit, $\bar m \to 0$.  
To see it explicitly, 
first rewrite identically (\ref{Mrenansatz1}) as 
\beq
M_1 (\ln (M_1/\Lam) )^{\frac{\gamma_0}{2b_0}} = \hat m
\label{M1rewritten}
\eeq
where for convenience we simply introduced 
the RG invariant scale $\Lam =
{\overline{\mu} e^{-{1\over{2b_0 g^2}}}}$
(at first RG order), and the {\em scale-invariant} mass
$\hat m \equiv \bar m (2b_0 \bar g^2)^{-\frac{\gamma_0}{2b_0}} $. For
 fixed $\bar g^2$, the chiral limit is $\hat m \to 0$ in 
(\ref{M1rewritten}). Now, (\ref{M1rewritten}) may be seen as a function
$\hat m(M_1)$, and requiring its inverse $M_1(\hat m)$ to be defined
on the whole physical domain $ 0 < \hat m < \infty$, and to match the 
ordinary perturbative asymptotic behavior for $ \hat m \to \infty$,
implies $M_1 (\hat m \to 0) \to \Lam $~\footnote{ 
There is a priori another possible solution of
(\ref{M1rewritten}) in the chiral limit: 
 $M_1 \to 0$ when $\hat m \to 0$. However, 
 it is easily seen that 
$M_1(\hat m)$ has branch points at $M_1 = \Lam$ and 
$M_1 = e^{-\gamma_0/2b_0} \Lam $. The branch giving the trivial solution
is only defined for $0 \le \vert \hat m \vert \le 
(\gamma_0/2b_0)^{\gamma_0/2b_0}
e^{-\gamma_0/2b_0} \Lam < \Lam$, and is therefore not compatible with the
asymptotic perturbative behavior of (\ref{Mrenansatz1}) for $\bar m \to
\infty$.}. 
 This property of (\ref{Mrenansatz1}) is in contrast with the  
``one-loop improved" expression of the   
mass (obtained by replacing
$M_1 \to \bar m$ in the denominator of (\ref{Mrenansatz1})), 
which would only give 
$M_1 \to 0$ for $\bar m \to 0$. 
In other words, while (\ref{Mbareansatz1})--(\ref{Mrenansatz1}) 
are perturbatively
compatible with the usual (one-loop) RG behaviour of the current mass,
 a specific choice
of boundary condition for the RG running has selected a non-trivial
``vacuum", by changing the behaviour of the whole perturbative
series in the chiral limit $\bar m \to 0$. 
Alternatively, one may derive the result
$M_1 = \Lam$ in the chiral limit as follows. 
From (\ref{M1rewritten}), defining
\beq
F = \ln [\frac{M_1}{\Lam}] \equiv \ln [\frac{\hat m}{\Lam}] -
\frac{\gamma_0}{2b_0} \ln F\;,  
\label{Fdef1}
\eeq
on the physical branch, $0 <\hat m < \infty$, 
the latter relation can be inverted as 
\beq
e^F  F^{\frac{\gamma_0}{2b_0}} = \frac{\hat m}{\Lam}\;.
\label{Finv1}
\eeq
 Now it is a
simple algebraic exercise to see that (\ref{Finv1}) gives 
$F(\hat m)$ as an expansion in powers of 
$(\hat m/\Lam)^{(2b_0/\gamma_0)}$, for small $\hat m$:
\beq
{\ds F (\hat m \to 0) \simeq  
(\frac{\hat m}{\Lam})^{2b_0\over \gamma_0}\; [ 1 -
{2b_0\over \gamma_0} 
(\frac{\hat m}{\Lam})^{2b_0\over \gamma_0} + \cdots \;] },
\label{Fexpm}
\eeq
which together with (\ref{M1rewritten}), i.e. 
$ M_1 = \hat m \: F^{-(\gamma_0/2b_0)} $, immediately gives
$M_1 \to \Lam $ for $\hat m \to 0$.
While the latter derivation may be unecessarily 
sophisticated at first order, it has the advantage of being the one
easily generalizable to higher order, where we shall also find
a non-zero mass in the chiral limit, $M = const \, . \,\Lam$, but 
 with a no longer trivial proportionnality constant. More importantly, 
it is also the most convenient procedure to analyse the chiral limit once 
having established an expression analoguous to 
 (\ref{Mrenansatz1}), but obtained directly from the 
variational perturbative (bare) expansion (as defined by the substitution
(\ref{substitution})), which we address next. Our aim is  
 to obtain a variational ``mass gap" where the 
 non-trivial chiral limit property of (\ref{Mrenansatz1}) 
is preserved, while
at the same time providing us with a systematically
 improvable ansatz, in accordance with the ``variationally improved
perturbation" principle. \\ 

Let us thus proceed  with the $x$ parameter expansion,
performing in (\ref{Mbareansatz1}) 
the substitution (\ref{substitution}).
Unfortunately, as announced in the last section, 
it is easily checked that the resulting expression
is no longer finite: after the introduction of the usual counterterms
in (\ref{zmdef}), some of the divergent terms in the denominator of 
eq.~(\ref{Mbareansatz1}) no longer cancel, except for $x \to 1$,
in which case one only obtains the trivial result $M_1 \to 0$. 
This is not much 
suprising since the precise all-order cancellations of divergences in
eq.~(\ref{Mbareansatz1}),
resulting from the above RG properties, have no a priori
reasons to be compatible with the peculiar modification of the basic
Lagrangian mass and coupling as implied by the 
substitution (\ref{substitution}). 
The way out
is to resum the $x$-generated series, denoted by $M_1(x)$, 
in a different manner~\cite{gn2}:
by analytical continuation in $x$ one can find an adequate 
integration contour, 
resumming exactly the series $M_1(x)$ 
in the $x \to 1$ limit. 
This is explained in full details
in Appendix B. 
The net result gives an ansatz (still 
at the first 
RG order approximation):
\beq
M_1  = {1\over{2 i \pi}} \oint {dv \over v} e^v
\; { v\; m_0 \over{f_0(v)^{\gamma_0
\over{2b_0}}} }\;,
\label{contour4}
\eeq
where 
$v \equiv q(1-x)$ has been introduced as an appropriate change of
variable to analyse the $x \to 1$, $q \to \infty$ limit of the 
$q^{th}$-order expansion of $M_1(x)$,
the contour is counterclockwise 
around the negative real axis, and we have 
defined 
\beq
f_0(v) \equiv 1- b_0 \; g^2_0 \;\Gamma[\frac{\epsilon}{2}]
\;(m_0 v)^{-\epsilon}\; (f_0)^{\epsilon \frac{\gamma_0}{2b_0}}
\label{fvdef}
\eeq
whose expression  
is directly dictated from the denominator of eq.~(\ref{Mbareansatz1})
(see Appendix B for details).
The crucial point is that (\ref{contour4}) 
is now finite to all orders,  
while giving a non-zero result. 
More precisely, 
after renormalization
eq.~(\ref{contour4})
becomes 
\beq
M_1 = {1\over{2 i \pi}} \oint dv e^{v} {\bar m \over {f(v)^{{\gamma_0
\over{2b_0}}}}} 
\label{contour5}
\eeq
where the 
renormalized function $f(v)$, related to $f_0(v)$ above as~\footnote{
The simple relation in (\ref{f0frel}), involving the counterterm
$Z_g$ only, is an accident
of the first RG order approximation. At higher orders the relation is much 
more involved~\cite{gn2}.}
\beq
f_0(1) = Z_g f(1),
\label{f0frel}
\eeq
 satisfies the (finite) recursion formula
\beq
f = 1 + 2b_0 \bar g^2 \ln({\bar m\;v
\over{\bar \mu}}\;f^{-\frac{\gamma_0}{2b_0}})\;
.
\label{f1def}
\eeq
Introducing 
a last convenient change of variable, 
\beq
F \equiv {f\over{2b_0 \bar g^2}}\;
\raisebox{-0.4cm}{~\shortstack{ $=$ \\ $ (v = 1)$}} 
\;\;\; \frac{1}{2b_0 g^2(M_1)}\;,
\label{Fdef}
\eeq
eq.~(\ref{contour5}) takes the form:
\bea
M_1 = 
{\Lam
\over{2
i \pi}} \oint dy e^{y/m^{''}} {1 \over {F^{{\gamma_0 \over{2b_0}}}}}
, 
\label{contour6}
\eea
where 
\beq  
{\ds 
m^{''} \equiv (\bar m/\Lam)\; (2b_0\bar g^2)^{-\gamma_0/2b_0}\; ;\;\;\;\; 
y \equiv m^{''} v\; } 
\label{msecdef}
\eeq
are just 
the conveniently rescaled, 
dimensionless (scale-invariant) ``mass" parameter 
and integration variable, respectively.
After those different manipulations, it is easy to see that 
one can use (\ref{Fdef1}) to
express $F$ in (\ref{contour6}) as an expansion
for small $y $, using (\ref{Fexpm}) with the substitution 
$(\hat m/\Lam) \to y$:
\beq
{\ds F (y \to 0) \simeq  y^{2b_0\over \gamma_0}\; ( 1 -
{2b_0\over \gamma_0} y^{2b_0\over \gamma_0} +
{\cal O}(y^{4b_0\over \gamma_0}) \;) }.
\label{Fexp}
\eeq
This is nothing but a manifestation of the dimensional
transmutation mechanism in the $m \to 0$ limit of QCD, 
although in a rather
unconventional form:
the function $F$, defined in (\ref{Fdef})
as a function of the running coupling becomes, for $m \to 0$, a function
of the (rescaled) mass only,
$y \equiv m'' v$. The original expression depending on
$\{\bar g^2, \bar m \}$ now depends on $\{ \Lam, m^{''} \}$ 
(where $m''$ is arbitrary), and we are 
ultimately interested in the chiral limit $m^{''} \to 0$.
From (\ref{Fexp}) it is clear that (\ref{contour6})
 has a simple pole
at $y \to 0$, with residue giving the announced
non-trivial solution, $M_1 = \Lam$.  

The previous construction therefore shows that summing the $x$ series
for  the
variational
(bare) expansion and renormalizing, obtaining  
the contour integral 
in (\ref{contour5}), is equivalent to performing 
the following steps:\\
-i) take the renormalized
RG solution of (\ref{runmass}), $M(\bar m)$,
 with the condition
(\ref{RGBC}); \\
-ii)  performing in $M(\bar m)$ 
the substitution $\bar m \to \bar m v$, 
and integrating the resulting expression 
around the cut negative real axis, with a specific weight 
$\oint (dv/v)\:e^v$ as in 
eq.~(\ref{contour5})~\footnote{Actually at the
next order, solving (\ref{runmass}) and taking the integral afterwards
is not strictly equivalent  
 to the derivation starting from 
the bare expressions. 
However one can still show~\cite{gn2}
 that the two derivations are simply related by 
a particular 
renormalization scheme change.}. Having this well-defined 
integral transform we can 
work directly with renormalized quantities and define formally 
an ansatz at a 
given order as an appropriate generalization of the 
integral in (\ref{contour5}), 
{\em provided} however that 
one does not encounter  
extra singularities in $v$. For simplicity
let us 
postpone the question of
the possible existence of extra singularities until
section 2.4, where this potential problem is addressed in details. \\ 

It is now straightforward to introduce the necessary non-logarithmic
perturbative corrections to the (purely RG) above results: 
this is consistently
done as
\bea
M^P_1 (m^{''}) =
{\Lam
\over{2
i \pi}} \oint dy e^{y/m^{''}} {1 \over {F^{{\gamma_0 \over{2b_0}}}}}
(1 +({2\over 3}){\gamma_0 \over{2b_0 F}} +{\cal O}({1\over F^2})\;)
\label{contour77}
\eea
without changing anything  
in the contour integral properties, except that 
(\ref{contour77})  no longer has a simple pole behavior at $y \to 0$. 
Eq.~(\ref{contour77}) is nothing but a specific integral over 
the pole mass: namely taking the integrant for $v=1$
and expanding for small $\bar g^2$ by using
relations (\ref{f1def}) and (\ref{Fdef}), one 
explicitly recovers the usual pole mass expression
with first order non-logarithmic perturbative correction. \\
One may now optimize
(\ref{contour77}) with respect to $m^{''}$, by performing 
(numerically) the $y$ integral. 
In fact, numerical integration is not mandatory: 
since what we are interested
in is the behavior 
for $m^{''} \to 0$, it is  
equivalent to look at the properties
of eq.~(\ref{contour77}) for $y \to 0$. 
An expansion of (\ref{contour77}) near the origin
is provided from Hankel's formula, 
\beq
\label{hankel}
\frac{1}{2i \pi} \oint dy e^{y/m^{''}}  y^\alpha  = 
\frac{(m^{''})^{1+\alpha}}{\Gamma[-\alpha]}\; ,
\eeq
where the different powers $\alpha$ 
resulting from the expansion at arbitrary 
order near the origin have the form $\alpha =-1 + p 
\frac{2b_0}{\gamma_0}$, with $p =-1,0,\cdots$ 
integer~\footnote{
Note that 
(\ref{contour77}) with (\ref{hankel}) has some similarities with  
a Laplace-Borel transform: indeed the analytic continuation
in $m$ is almost equivalent (for $m \to 0$) 
to a continuation in the coupling $g$.
The precise connection, and the link with the 
renormalon singularities~\cite{renormalons}
associated with the usual Borel transform
will be discussed in more details
elsewhere~\cite{jlk}. We shall however come back in the next sections on
the structure of singularities of the integral (\ref{contour77})
encountered
in our framework.}.\\

We stress that
$m^{''}$ (equivalently, $\bar m$)
is meant to be an arbitrary but implicitly {\it small} parameter:
even if we ultimately
seek for optimal values $m^{''}_{opt}$ as best approximations
to the limit $m^{''} \to 0$, at intermediate steps
$m^{''} \neq 0$ (i.e $\bar m \neq 0$) implies that the axial vector
current Ward-Takahashi identities are explicitly broken.
Eventually a small, physically acceptable explicit breaking 
(like the PCAC hypothesis~\cite{LET}) may be considered,  
where $m^{''}_{opt}$ ($\bar m_{opt}$) could be tentatively 
interpreted
as the actual physical quark masses. But that is largely 
in contradiction with the basic principle adopted here, according to
which $m''$ itself has no physical meaning, whereas only 
the optimal
value of $M^P_1(m'')$ in (\ref{contour77}) has.
Indeed, in the (ideal)
situation where the optima would be really flat, 
$m''$ would obviously not be well determined. 
It is therefore much more preferable to 
find a mean, eventually an approximate one, 
 to reach the exact limit $m^{''} \to 0$. 
The chiral Ward identities are then recovered and
we can define the mass gap
strictly in the chiral symmetric limit.
We shall see in section 4 
how to implement consistently an explicit symmetry
breaking, physical mass term, independent of the arbitrary 
variational mass parameter $m''$.
\subsection{Second order dynamical mass ansatz}

At the next RG order, 
the previous qualitative picture remains essentially unchanged, except
that the derivation is somewhat more involved. 
The solution of (\ref{runmass}) with 
(\ref{RGBC}), using 
the two-loop RG coefficients~\cite{b0,b1} given in Appendix A, is 
\beq
M_2 =  \bar m \;\;
\ds{f^{-\frac{ \gamma_0}{2b_0}}\;\; \Bigl[\frac{ 1 +\frac{b_1}{b_0}
\bar g^2 f^{-1}}{ 1+\frac{b_1}{b_0}\bar g^2} \Bigr]^{ -\frac{\gamma_1}{
2 b_1}
+\frac{\gamma_0}{2 b_0}   } }\;
\label{MRG2}
\eeq
where
$f \equiv \bar g^2/g^2(M_2)$ satisfies 
\beq
f = \ds{ 1 +2b_0 \bar g^2 \ln \frac{M_2}{\bar \mu }
+\frac{b_1}{b_0} \bar g^2
\ln \Bigl[\frac{ 1 +\frac{b_1}{b_0} \bar g^2 f^{-1}}{
 1 +\frac{b_1}{b_0} \bar g^2 }\;f\;\Bigr] }\; ;
\label{f2def}
\eeq
(note in (\ref{MRG2})--(\ref{f2def}) the recursivity in both $f$ and $M_2$), 
and the non-logarithmic perturbative corrections are easily 
included as~\footnote{In (\ref{f2def}) the recursivity
operates on $M_2$, 
which by definition only depends on the RG evolution, therefore 
to be distinguished from 
the (pole) mass expression $M^P_2$ in (\ref{Mpole}).}
\beq
M^P_2 \equiv M_2 \;\Bigl(1 +{2\over 3}\gamma_0 {\bar g^2\over f}
+{K \over{(4 \pi^2)^2}}{\bar g^4\over f^2}+{\cal O}(g^6)\;\Bigr)
\label{Mpole}
\eeq
where 
the complicated  
two-loop coefficient $K$ was calculated
exactly in ref.~\cite{Gray} and is given explicitly
in Appendix C.  
Eq. (\ref{Mpole}) defines the pole mass including two-loop non-logarithmic
corrections, and can be easily 
shown in addition to resum the leading
{\it and} next-to-leading logarithmic dependence in $\bar m$ to all 
orders (see Appendix A). 
The contour integral generalization of (\ref{contour77}) is obtained,
as explained before,  
after making the substitution $\bar m \to \bar m v$ in 
(\ref{MRG2})--(\ref{Mpole}), as 
\beq
{ M^P_2 (m^{''}, a)\over \Lam}
 = {2^{-C} a\over{2 i \pi}} \oint dy {e^{y/m^{''}}
\over{F(y)^A [C + F(y)]^B}} {(1 +{{\cal M}_{1}(a)\over{F(y)}}
+{{\cal M}_{2}(a)\over{F(y)^2}} )},
\label{contour7}
\eeq
where $\Lam$ is now the RG invariant basic scale
at two-loop order, in 
$\MS$~\cite{lambdabar}~\footnote{With our prescritions  
one should keep in mind that $\Lam$ depends on $n_f$, 
as usual~\cite{Marciano}.}:
\beq
{\ds 
\Lam \equiv \overline{\mu} 
e^{-{1\over{2b_0 \bar g^2}}} (b_0 \bar g^2)^{-{b_1
\over {2b^2_0}}} (1 +{b_1\over b_0} \bar g^2)^{{b_1
\over {2b^2_0}}} }\; , 
\label{Lam}
\eeq 
and the dimensionless (scale-invariant) 
arbitrary  mass parameter reads 
\beq
m''\equiv  \ds{(\frac{\bar m}{ \Lam}) \;
2^{C}\;[2b_0 \bar g^2]^{-\gamma_0/(2b_0)}
\;\left[1+\frac{b_1}{b_0}\bar g^2\right]^B}
\; .
\label{msec2def}
\eeq
In (\ref{contour7}) $F$ now satisfies the recursive relation 
\beq
F(y) \equiv \ln [y] -A \ln [F(y)] -(B-C) \ln [C +F(y)],
\label{Frec2}
\eeq
where A, B, C are expressed in term of RG coefficients as
\beq
{A = {\gamma_1\over{2 b_1}}},\;\;\; {B = {\gamma_0\over{2 b_0}}
-{\gamma_1\over{2 b_1}}},\;\;\; {C = {b_1\over{2 b^2_0}}}\; ,
\label{defABC}
\eeq
explicitly given in Appendix A. \\

In (\ref{contour7}) we have also introduced an extra parameter $a$, taking  
into account changes in the arbitrary renormalization scale, 
according to $\bar \mu \to a\; \bar \mu$. Indeed, 
even at the first RG order there are
infinitely many ways of introducing the non-logarithmic perturbative 
corrections to (\ref{contour6}): for
instance, we might have introduced the ${\cal O}(1/F)$ terms 
via another definition of $F$: $F^{'} = F(1+{\cal O}(1/F))$
in place of (\ref{Fdef1}),  
up to second order terms. This
non-uniqueness is basically what is parameterized by $a$, 
and accordingly the 
perturbative coefficients ${\cal M}_{i}$ in (\ref{contour7}) have a 
logarithmic dependence in $a$, simply  
dictated order by order from the requirement that (\ref{contour7})
only differs from the original $\MS$ scheme expression by 
higher order terms:
\beq
{\cal M}_{1}(a) = {{\gamma_0\over{2 b_0}}({2 \over 3} -\ln ~ a)},
\eeq
\beq
{\cal M}_{2}(a) = {{1\over{(2 b_0)^2}}({K\over{(4 \pi^2)^2}} +\gamma_0 (
{\gamma_0 \over 2} +b_0) \ln^2~ a
-({4\over 3} \gamma_0 b_0 -{\gamma^2_0\over 3} +\gamma_1)
 \ln ~ a)}\; .
\label{M1M2}
\eeq
Since we shall however 
continue expression (\ref{contour7}) to the (non-perturbative)
region of $m'' \to 0$, in the manner described in the next sections, the
$a$-dependence will eventually exhibits non-trivial extrema, 
and it is thus sensible to 
optimize the result with respect to $a$~\footnote{
This procedure indeed gave very good results~\cite{gn2}
in the GN model, where in particular for low values of $N$ the
optimal values $a_{opt}$ are quite different from 1.}, 
since the unknown exact result
would not depend on the arbitrary renormalization scale
$\mu$.
Note that $a \neq 1$ corresponds to a change of renormalisation {\em scale}
 only, while 
keeping the renormalization {\em scheme} fixed, i.e. 
$\bar g^2 = g^2(\bar \mu)$ (equivalently
$\Lam $) and $\bar m = m(\bar \mu)$ fixed 
(to $\MS$).
Actually the situation is slightly more complicated, 
since  at second perturbative order 
there are other possible changes of renormalization prescriptions 
than a simple change of scale, which do 
affect expression (\ref{contour7}) 
and should therefore be
taken into account in principle. 
This turns out to be an important aspect of 
our analysis and will be addressed in details in the next section. \\ 

For small $y$, $F(y)$ from (\ref{Frec2}) has the expansion: 
\beq
F( y \to 0) \simeq A (u -{B \over C} u^2 +{\cal O}(u^3));\;\;\; u \equiv
A^{-1} C^{{B-C}\over A}\;y^{1\over A}\; ,
\label{Fexp2}
\eeq
again implying that (\ref{contour7}) would 
give a simple pole at $y \to 0$  
for vanishing perturbative correction terms, ${\cal M}_1 ={\cal M}_2 = 0$,
with residue $(2C)^{-C} $.
Note that all the results of section 2.2 are
consistently recovered by taking $b_1 = \gamma_1 = 0$
and neglecting the non-logarithmic 
perturbative corrections of ${\cal O}(g^4)$
in the different expressions above, (\ref{MRG2})--(\ref{M1M2}). 

It is still possible to generalize to the next RG order the 
previous derivation,
since in QCD the three-loop coefficients $b_2$, $\gamma_2$
are 
known~\cite{b2}, although the construction becomes quite cumbersome.
In fact such a generalization is not worth doing, since 
one can in fact choose
a renormalization scheme in which $
b_2$ and $\gamma_2$ are set to zero, as well as all 
higher order coefficients. This is examined in more details in the next 
section. 

\subsection{Renormalization scheme changes and generalized ansatz}
As it turns out, the simple picture emerging from the last section is 
unfortunately questionable as one realizes 
that, strictly speaking,  
formula (\ref{contour7}) has  
extra singularities in the $y$ plane, in addition 
to the cut on the negative real $y$ axis,
implicit in its derivation and in (\ref{hankel}).   
After all, this specific contour 
was suggested by the known properties of 
the GN model, and it is not surprising 
if the analytic continuation 
in $m$ properties are more complicated in QCD. 
To begin, as is clear from (\ref{contour7}) there is an 
extra branch point at $F= -C = -b_1/(2b^2_0)$, which was not present in
the first order ansatz (\ref{contour77}), and simply corresponds to
the first non-trivial fixed point of $\beta(g)$ located at 
$\bar g^2 = -b_0/b_1$. Fortunately it is harmless in the present 
QCD case, 
since 
$b_1 > 0$ implies that it is located along
the already cutted $F < 0$ real axis~\footnote{In the GN model case the
similar cut was more troublesome, since $b^{GN}_1 < 0$,  
and we used a specific Pad\'e approximant construction~\cite{gn2}
avoiding the cut. In a different context, 
this non-trivial fixed point has also 
been studied recently~\cite{PerRaf} in 
connections with renormalon properties.}. \\
Less trivially, 
the zeros of $dy/dF$ 
also give from (\ref{Frec2}) extra
 branch cuts in the $y$ plane, 
starting at the (complex conjugate) points  
\beq
 y_{cut}(\gamma_1) = e^{F(\gamma_1)} F^A(\gamma_1) 
[C +F(\gamma_1)]^{(B-C)} \; ;
\label{ycut}
\eeq
where
\beq
{\ds F(\gamma_1) = \frac{1}{2}(-\frac{\gamma_0}{2b_0})[1 \pm 
(1-4\frac{\gamma_1}{\gamma^2_0}
)^{1/2}]\; }. 
\label{Fcut}
\eeq 
However, as can be seen the actual position of those branch points
do depend on the scheme via the second coefficient of the anomalous 
mass dimension, $\gamma_1$ (the precise RS dependence
of the latter 
is given in eqs.~(\ref{RSchange}), (\ref{RSgamma}) of Appendix A). 
In the original $\MS$-scheme, $F(\gamma^{\MS}_1)$
gives
extra cuts starting
at $Re[y_{cut}] \simeq$ 0.34 (0.24), $Im[y_{cut}] \simeq \pm $ 0.74 
($\pm$0.76)
for $n_f =2$ ($n_f =3$) respectively.
Note that when looking at the limit of interest,  
$y \simeq 0$, using the expansion in (\ref{Fexp2}),  
one never sees
those extra cuts: at any {\em finite} order 
the series expansion only has the cut on the negative real axis
according to eqs.~(\ref{hankel}) and (\ref{Fexp2}). 
These singularities may 
after all be an artifact of our extrapolation to very small $m''$ of a
``perturbative" (although resummed) relation, eq. (\ref{Frec2}). 
Anyhow,  
the point is that using the expansion near the origin  
is invalidated if there are extra singularities lying in the 
way with $Re[y_{cut}] 
>0$, since 
it would lead to an ambiguity of ${\cal O}(\exp({\rm Re}[y_{cut}]/m''))$
for $m'' \to 0$, in the determination of the 
integral (\ref{contour7})~\footnote{This is
quite similar to the renormalon ambiguities~\cite{renormalons}. 
Although the {\em usual} renormalons are not explicitly seen 
here due to the appearance of a mass gap (i.e., within the mass gap
ansatz (\ref{contour7}), by construction 
there is no
integration over the Landau pole region), 
the way in which the singularities
in $y$ 
appear, namely in a resummed expression relating a ``reference" scale
$M_{dyn} \simeq \Lam $ to an infrared scale $m'' \simeq 0$, 
may be viewed as reminiscent
from the renormalons.  
An essential difference is that in the present case those
singularities occur in the analytic continuation of the mass parameter
rather than the coupling, and 
that 
it is possible to move 
those singularities away by appropriate RS change,
as is discussed below.}. \\

The way out is thus clear: if there exists values of $\gamma_1$
which move those extra cuts away (or which are such that they 
start at $Re[y_{cut}] = 0$),  
the expansion around the origin in $y$ is legitimate. 
Defining 
\beq
\gamma^{'}_1 \equiv \gamma_1 +\Delta \gamma_1\; ,
\label{dga1}
\eeq
one easily finds that 
$Re[y_{cut}] \simeq 0$ for $\Delta\gamma_1 \simeq$ 0.00267 (0.00437) for 
$n_f =$ 3 ($n_f =$ 2) respectively~\footnote{Actually $Re[y_{cut}]$ is a 
(semi)-periodic function of $\Delta\gamma_1$. We only consider
solutions of $Re[y_{cut}(\Delta\gamma_1)] \simeq  0$ 
nearest to the
original $\MS$ value ($\Delta\gamma_1 \equiv 0$) for $n_f =2,3$.}.  
Therefore we can adjust a correct $\gamma^{'}_1$ by performing 
(perturbative) RS changes. 
As explicit from eqs. (\ref{RSchange}), (\ref{RSgamma}) of Appendix A, 
this can
be done either by a first order change in $g \to g^{'}$ {\em or} 
$ m \to m^{'}$ (or both).
One thus considers the most general RS change (but restricted to the 
second perturbative order, which is sufficient for our purpose) 
and optimize with respect
to this new arbitrariness. (As a side remark, we mention that
one would also have $Re[y_{cut}] < 0$ with $\gamma_1 \equiv 
\gamma^{\MS}_1$, 
for $n_f \geq 5$). Note that the removing of the
unwanted singularities is only possible at the two-loop RG order
ansatz, eq.~(\ref{contour7}), due to the first occurence of RS
arbitrariness  
at two-loop order via $\gamma_1$, in MS schemes. More
precisely, for  
the first order ansatz, eq.~(\ref{contour77}), an extra singularity
occurs  now with  
$Re[y_{cut}] = Re[\exp[-\gamma_0/2b_0]\;(-\gamma_0/2b_0)^
{\gamma_0/2b_0}] > 0$, which accordingly cannot be removed by a 
RS change. \\

We obtain
after simple algebra the generalized dynamical mass ansatz 
in the new (primed) scheme, in terms of the arbitrary RS change
parameters $A_1$, $A_2$, $B_1$, $B_2$ defined in Appendix A):
\beq
{ M^P_2 (a, m^{'''})\over \Lam^{'}}
 = {2^{-C} a\over{2 i \pi}} \oint dy {e^{y/m^{'''}}
\over{(F^{'}(y))^{A^{'}} [C + F^{'}(y)]^{B^{'}}}} 
{(1 +{{\cal M}^{'}_{1}(a)\over{F^{'}(y)}}
+{{\cal M}^{'}_{2}(a)\over{F^{'2}(y)}} )},
\label{contour8}
\eeq
with
\bea 
& {\cal M}^{'}_{1}(a, B_1) = &
{\cal M}_{1}(a) -\frac{B_1}{2b_0}\; ;  \nn \\
& {\cal M}^{'}_{2}(a, A_1, B_1, B_2) = & 
{\cal M}_{2}(a) -(A_1+B_1) 
\frac{{\cal M}^{'}_{1}(a)}{2b_0} -\frac{(B_2-\gamma_0 B_1)}{4b^2_0}
\label{M1M2RS}
\eea
and 
\beq
\Lam^{'} = exp[\frac{A_1}{2b_0}]\; \Lam \; 
\label{LamRS}
\eeq
(where relation (\ref{LamRS}) is exact to all orders~\cite{CelGon})
and $F^{'}$ and $m^{'''}$ have specific expansions in terms
of the original $\MS$ quantities $F$ and $m^{''}$
that we will not need
explicitly here. 
We also impose a further RS choice,
\beq 
b^{'}_2 \equiv 0\; ;\;\;\; 
\gamma^{'}_2 \equiv 0\; ,
\label{b2andg2}
\eeq
which, 
according to relations (\ref{RSbeta})--(\ref{RSgamma})
in Appendix A, fixes $A_2$
and $B_2$ uniquely in terms of $B_1$ and $\Delta\gamma_1$~\footnote{
Although e.g. $A_2$ does not appear explicitly
in (\ref{contour8}), we nevertheless need a prescrition to fix it,
since it appears at the second order general RS change.}
and guarantees that the definition of $\Lam$
in (\ref{Lam}) is unaffected, apart obviously from (\ref{LamRS}).
In what follows we express all results in terms of the original
$\Lam$ scale.

\subsection{Pad\'e approximants and numerical results}
 
It is worth emphasizing at this point that our purpose is not
to find a particular RS choice, which would ``best fit"
the expected order-of-magnitude result for the (dynamical) constituent mass 
(or similarly for 
$F_\pi$ and $m\qq$ considered in the next sections). Indeed, it is very
likely that with so much RS freedom at disposal, one could make
eq.~(\ref{contour8})
 fit almost whatever values one wishes. In contrast, what 
we are seeking is the {\em flattest} region in the arbitrary RS parameter
space, in the ``Principle of Minimal Sensitivity" (PMS)~\footnote{
For an extended discussion of the PMS motivations, we refer to the
original Stevenson's paper, ref.~\cite{pms}.} sense. 
One soon realizes however that 
our extension of the PMS
defines a rather complicated optimization  
problem: one has in principle to find the flattest
possible extrema of (\ref{contour8}), in the three independent parameter 
space $\{m^{''}, a, B_1 \}$, where in addition $\Delta\gamma_1$ 
is constrained to give good analyticity behavior of (\ref{contour8}).
 
Fortunately, one can study this problem within some approximations, 
which we believe are legitimate. 
As above explained the ansatz in 
(\ref{contour7}) (or (\ref{contour8})) is already optimal with respect
to $m^{''}$ at $m'' =0$, by construction,
for vanishing pertubative non-logarithmic 
corrections ${\cal M}_{i} =0$, and in this case the optimal result
is the simple pole residue.
Due to the non-logarithmic, {\em purely perturbative}
corrections (which are at present 
only known to second order for the mass), this simple picture
is lost, but accordingly
one may assume that the resulting
expansion for small $m''$ is ``as close as possible'' to an optimum 
(as would be the case if the sequence of approximations obtained by
considering increasing orders of the $x$ expansion could be proved to
converge, like in one-dimensional models~\cite{JONES}). 
Accordingly, we will 
define the $m'' \to 0$ limit of (\ref{contour8}) by a relatively
crude but standard approximation of those perturbative
corrections, rather than performing a numerical optimization
with respect to $m''$. 
Indeed, as discussed at the end of
section 2.2, it is in addition  
physically motivated to reach $m'' \to 0$,
since the axial vector current
Ward-Takahashi identities are recovered and
a non-zero result signals the spontaneous chiral symmetry
breakdown. \\

The approximation we are looking for is 
certainly not unique: given the ansatz (\ref{contour8}), one 
may construct different approximant forms leading to  
a finite limit
for $m'' \to 0$. 
We shall demonstrate
the feasibility of our program in the simplest 
realization: 
since the resulting expression will anyhow
be optimized with respect to the RS dependence
(entering any such approximants via the 
RS dependence of ${\cal M}_{i}$), we assume that it largely 
takes into
account this non-uniqueness due to higher order uncertainties, 
in the standard
PMS sense. The latter assumption is also supported by the results in the
GN model~\cite{gn2}, where several different approximants were tried and 
compared with the known exact results. 

Pad\'e approximants are generally known to greatly improve perturbative
results~\cite{pade}
and in most cases have the effect of smoothing the RS dependence.  
We shall consider 
the following Pad\'e approximant for the purely perturbative
part of the integrant in (\ref{contour8}):
\beq
Pad\acute{e}\;(F) 
\equiv {F +\lambda(a, B_1)\over{F +\rho(a, B_1)}} 
\raisebox{-0.3cm}{~\shortstack{ $\to $ \\ $ F \to 0 $}}
\;\;\;{\lambda(a) \over{\rho(a)}},
\label{padeM}
\eeq
which by construction restitutes a simple pole for $F \to 0$ in 
(\ref{contour8}). Matching its 
perturbative expansion 
for $F \to \infty$ to the one in (\ref{contour8}) one obtains  
\beq
M^{Pad\acute{e}}_2(a,\Delta\gamma_1,B1) 
\raisebox{-0.4cm}{~\shortstack{ $=$ \\ $ m'' \to 0 $}}
\Lam\;
(2C)^{-C} \;a\; exp\{\frac{A_1}{2b_0}\}\;\left[
1 -{{\cal M}^2_{1}(a, \Delta\gamma_1, B_1)\over{{\cal M}_{2}(a, 
\Delta\gamma_1,B_1)}}\right]\;. 
\label{Mpade}
\eeq
When defining (\ref{padeM})--(\ref{Mpade})
 one should be careful not to introduce  
new poles in the $F >0$ ($y >0$) plane. This simply implies that
there are some constraints on the possible values of $(a, B_1)$ which,
if we are lucky, do not invalidate their optimal values obtained from 
minimization. \\

We have performed a rather systematic study of the possible extrema
of the Pad\'e approximant formula (\ref{Mpade}) for arbitrary 
$a$, $B_1$, with $\Delta\gamma_1$ 
fixed such that the extra cuts start at $ Re[y] \simeq 0 $.
We did find an optimal region
with respect
to $a$, in the sense that it minimizes the
second derivative at the maximum~\footnote{Note that 
the second derivative 
at the extrema points is a good  
quantitative estimate of the flatness criteria even in a multi-dimensional
parameter space: 
the intrinsic curvature at an extremum 
of the hypersurface as  
defined from e.g. the approximant (\ref{Mpade}), 
is proportional to the
product of its second derivatives with respect to the 
different parameters.}.
The region of the
parameter space in the vicinity of that extremum is illustrated 
in Fig. 1,
as function of $a$ for different values of $B_1$. 
The plateau region has been determined more
accurately using a numerical steepest descent method, requiring
a minimal curvature. (For instance for $n_f =2$ 
it corresponds to $a_{opt} \simeq 1.42$, $B_{1,opt} \simeq 0.12$).
Our results are also summarized in Table 1 (see section 5). 
The optimal values of the parameters $a$, $B_1$, 
$\Delta\gamma_1$  are consistent 
with the further requirement
that (\ref{padeM}) has no poles at $F > 0$, as it should. 
Explicitly we obtain 
\beq
M^{Pad\acute{e}}_2(opt) \simeq 2.97\;\Lam(2)\;\;\;\;
(2.85\;\Lam(3)\;)\;,
\label{Mnum}
\eeq
for $n_f=2$ ($n_f=3$) respectively. \\
We shall see in the next section that we can relate
the pion decay constant $F_\pi$ to $\Lam(2)$ or $\Lam(3)$,
from which one may eliminate $\Lam$ to obtain an evaluation of
$M^{Pad\acute{e}}_2$ in the chiral limit. 
Anticipating on these determination of $\Lam$,   
eq.~(\ref{Mnum}) with $F_\pi \simeq $ 92 GeV
gives (still in the
pure chiral limit)
\beq
M^{Pad\acute{e}}_2(opt) \simeq  500\;(447)  MeV 
\eeq
for $n_f=2$ ($n_f=3$). \\
Finally, for a useful comparison, we also give the value of 
(\ref{Mpade}) in the $\MS$ scheme (which may be thus considered as
optimized with respect to $m''$ but in a fixed RS):
\beq
M^{Pad\acute{e}}_2(\MS) \simeq 0.99\;\Lam(2)\;\;\;\;
(1.00\;\Lam(3)\;)\;.
\label{MnumMS}
\eeq
Note however that according to the previous discussion, 
the result in (\ref{MnumMS})
are a priori
plagued by an ambiguity associated to the bad-placed extra
singularities in the original $\MS$ scheme. This may however 
give a qualitative idea of the effects of the
above optimization with respect to the RS choice.

\section{Generalization to Composite operators: $F_\pi/\Lam$}
\setcounter{equation}{0}
\subsection{RG-invariant ansatz for a composite operator}
We shall now derive an ansatz 
similar to eqs.~(\ref{contour7}),
(\ref{contour8})
 for the pion decay constant $F_\pi$. The main idea is to
do perturbation theory around the {\em same} RG evolution solution 
with the non-trivial fixed point, as specified by the function $F$ in
(\ref{Frec2}),  
with perturbative corrections specific to $F_\pi$ obviously.
One should first identify $F_\pi$ from a relation where the previous
construction can be best generalized. A formal definition 
which suits all our purposes is the well-known 
low-energy expansion of the axial vector--axial vector
two point correlation function~\cite{GaLeut,derafael}
\beq
i\;\int d^4q e^{iq.x} <0 \vert T\;A^i_\mu(x) A^k_\nu(0) \vert 0> \simeq
\delta^{ik} g_{\mu \nu} F^2_\pi +{\cal O}(p_\mu p_\nu)\;.
\label{Fpidef}
\eeq
In (\ref{Fpidef}) T is the time-ordered product and
$A^i_\mu$ the axial quark current, $A^i_\mu \equiv 
 (\bar q \gamma_\mu
\gamma_5\lambda^i q)/2$, 
where the $\lambda^i$'s are Gell-Mann $SU(3)$ matrices
or Pauli matrices for $n_f =3$, $n_f=2$ respectively.  
The non-vanishing of expression (\ref{Fpidef}) implies 
CSB: in other words $F_\pi$ is to be considered as an order
parameter~\cite{Sternetal,Stern,Leut1}. 
Since $F_\pi$ is expressed in terms of a gauge-invariant
and RG invariant composite operator in (\ref{Fpidef}),  one can 
apply a rather 
straightforward generalization of our ansatz exploiting 
in particular the RG 
resummation properties.
The perturbative expression of (\ref{Fpidef}) for $m \neq 0$ is known
to the three-loop order~\cite{Abdel,Avdeev}, 
and is given explicitly in $\MS$ scheme 
in Appendix C. \\ 
With a little
bit of insight, a generalization of 
(\ref{MRG2})--(\ref{Mpole}) to a composite operator 
${\cal O}^n$ of naive mass 
dimension $n$ (depending only on $g$ and $m$), can be written as 
\beq
{\cal O}^n \simeq \frac{2b_0 \; m^n }{F^{n A-1} [C+F]^{n B}}\;
\delta \;(1 +\frac{\alpha}{F} +\frac{\beta}{F^2}+\cdots) \; ,
\label{Onansatz1}
\eeq
in terms of $F$ as defined by eq.~(\ref{Frec2})
 (for $y/m^{''} \equiv v =1$), and where 
$\delta$, $\alpha$, $\beta$
are fixed by matching the perturbative
expansion in a way to be specified next. In practice, we shall
only consider (\ref{Onansatz1}) for 
the relevant cases of $n=2$ and $n=4$, corresponding to
$F^2_\pi$ and $m \qq$ respectively.\\ 
Formula (\ref{Onansatz1}) as it stands is not yet our final ansatz
and necessitates some comments. Apart from the trivial powers of $n$ 
dictated by dimensional analysis, a rather obvious 
difference with the mass formula is that  
the composite operator being not a Lagrangian term, 
its perturbative expression 
starts at the one-loop, but zero$^{th}$ order 
in $g^2$. 
To accomodate this fact with the correct RG properties, 
the expansion of (\ref{Onansatz1}) starts 
with a $1/g^2$, due to the extra $2b_0 F$ factor. 
The $1/g^2$
first order term anyhow cancels
after a necessary subtraction which we discuss 
now~\footnote{Note that the form 
and properties of (\ref{Onansatz1}) can
alternatively be rigorously derived by following
a construction starting from bare quantities, 
similar to the one explained in \cite{gn2} for the
vacuum energy of the GN model.}. 

A more essential difference with the mass expression  
is that (\ref{Fpidef}) 
is not finite even after mass and coupling 
constant renormalization. One can define a new, finite quantity after 
an adequate subtraction.
In the present case, since our construction starts from perturbative
expressions, this turns out to be nothing but the usual way  
of renormalizing a composite operator. 
Accordingly the prescrition is (perturbatively)
well-defined~\cite{Collins} and unambiguous for a given 
RS choice.
What is lost, however, 
is a part of the predictive power: unlike the mass case,
a consistent treatment
of the subtracted terms (i.e. respecting RG invariance) 
implies that the unambiguous determination of the $1/F^n$ perturbative
terms in the final 
ansatz necessitates information on the $(n+1)$ order
of perturbation theory.\\ 
In the two cases 
here considered, $F^2_\pi$ and $\bar m \qq $ in the
next section, this renormalization actually reduces to a  
simple subtraction of the 
operator $[m^n\;1]$~\footnote{
In the general case~\cite{ZuberKStern}, the 
renormalization of dim= 4 (gauge-invariant) operators 
involves a mixing matrix for 
the operators $m \qq $, $m^4\;1$, {\em and} $G^{\mu\nu}G_{\mu\nu}$.
However, the entry for $m \qq $ only involves $m^4\;1$ and 
$m \qq $ itself~\cite{Collins}. As for $F^2_\pi$, the only 
gauge-invariant operator of dim =2 is $m^2 1$.}. 
This implies, from 
RG invariance, residual finite subtraction terms, affecting 
the perturbative 
correction terms $\alpha$ and $\beta$  
in (\ref{Onansatz1}). 
Accordingly (\ref{Onansatz1}) is modifed 
to \beq
{\cal O}^n \simeq \frac{2b_0 \bar m^n }{F^{n A-1} [C+F]^{n B}}\;
\delta\;(1 +\frac{\alpha^{'}}{F} +\frac{\beta^{'}}{F^2}+\cdots)
-\frac{\bar m^n}{\bar g^2}\;H(\bar g^2)\; ,
\label{Onsub}
\eeq
where it is easily shown, using RG properties, 
that the finite subtraction function $H(\bar g^2) \equiv 
\sum^{\infty}_{i=0} H_i \bar g^{2i}$ 
is determined perturbatively order by order (in the $\MS$
scheme) from 
\beq
[n \: \gamma_m(\bar g) +{2 \over{\bar g}} \beta(\bar g) -
\beta(\bar g) {\partial \over{\partial \bar g}}]\;
H(\bar g^2) =
{\bar g \over 2}\; {\partial \;c_1(\bar g^2) \over{\partial \bar g}}\; .
\label{defsubtrac}
\eeq
In (\ref{defsubtrac}) $c_1(\bar g^2)$
 is given by the residue of the $1/\eps$ term in the perturbative series 
expansion for the relevant quantities, as given in (\ref{fpi2pert0}) 
and (\ref{mqqpert0}) 
respectively for $F^2_\pi$ and $m \qq$. 
The consistency of 
our formalism is checked by noting that the expansion of (\ref{Onsub})
in powers of $\bar g^2$
do reproduce correctly the LL and NLL dependence in $\bar m$ of the
perturbative expansion of the composite operator to all orders,
as well as the perturbative non-logarithmic terms explicitly displayed
in eqs.~(\ref{fpi2pert0}) and (\ref{mqqpert0}). 

\subsection{Generalized scheme ansatz for $F_\pi$}
From the results of the last section one can now write an ansatz 
for $F_\pi$, also taking into
account the most general RS dependence to second order
in a straightforward manner. 
After some algebra:
\bea
& \ds{
{F^2_\pi \over{\Lam^2}} = \exp[\frac{2A_1}{2b_0}]\; 2b_0\;
2^{-2 C} {a^2\over{2 i \pi}} \oint {dy\over y}\; y^2 {e^{y/m^{'''}}}
{\frac{1}{F^{'\;2 A^{'}-1} [C + F^{'}]^{\;2 B^{'}} }} 
}
\; \times \nn \\
& 
{\ds 
 \delta_{\pi}
 {\left(1 +{\alpha^{'}_{\pi}(a)\over{F^{'}}}+{\beta^{'}_{\pi}(a)
\over{F^{'2}}}
\;+\cdots \right)} 
}
\label{Fpiansatz}
\eea
where 
\beq
\delta_{\pi} = {N_c\over {2\pi^2}}{1\over{\gamma_0 -b_0}},
\label{deltapi}
\eeq
\beq
\alpha^{'}_{\pi}(a, \Delta\gamma_1) = 
\frac{1}{2b_0} \left[{5\over 6}(\gamma_0 -b_0) -2\pi^2
(\gamma^{'}_1 -b_1) -2(\gamma_0 -b_0) \ln\:a \right]\; ;
\label{alphapi}
\eeq

\bea
& \beta^{'}_{\pi}(a, \Delta \gamma_1, B_1) = \frac{1}{(2b_0)^2} 
\left[ f^{(2)}_\pi (\Delta\gamma_1) +2\gamma_0(
\gamma_0-b_0)
\ln^2\:a +\frac{(\gamma_0-b_0)}{6 \pi^2} \ln\:a \right] 
\nn \\ & -\frac{2 B_1}{2b_0} \alpha_\pi(a, \Delta\gamma_1)  
 +\frac{B_1}{(2b_0)^2}(\frac{11}{3}(\gamma_0-b_0)-4 
\pi^2(\gamma_1 -b_1))\; ,
\label{betapi}
\eea
where $f^{(2)}_\pi$ is a complicated expression given explicitly 
in Appendix C. \\
An important point is that
the subtraction function $H(\bar g^2)$ in (\ref{Onsub}) gives {\em no}
extra contributions to the $y$ integral~\cite{gn2}: 
after introducing the contour integration
as described in section 2 and Appendix B, the purely perturbative 
subtraction 
gives an analytic
function of $y$. Obviously however, 
the actual values of the perturbative coefficients 
in (\ref{alphapi}), (\ref{betapi}) are
affected
by the subtraction. As 
previously mentioned, 
the unambiguous determination of $f^{(2)}_\pi$ in
(\ref{betapi}) involves, in a given (minimal) RS, 
the knowledge of the
three-loop coefficient of   
$1/\epsilon$, obtained from the perturbative 
expression eq.(\ref{fpi2pert0}) of Appendix C.   
Note also that the dependence upon $\Delta\gamma_1$ in (\ref{alphapi})
is only through $\gamma^{\MS}_1 \to \gamma^{'}_1$,  
and the dependence on $B_1$ only appears in the second order term,
$\beta^{'}_{\pi}$, in contrast with the mass case. As well there is 
no dependence upon $B_2$ or $A_2$ at this (second) order. 
All these properties 
are of course consequences of 
the perturbative $F^2_\pi$
expression starting at one-loop but zero$^{th}$ $g^2$ order.\\

The discussion of the previous section on the
analyticity domain of (\ref{Fpiansatz}) with respect
to $\Delta\gamma_1$ is identical, 
since the branch cuts are determined by the very same 
relation defining $F$, eq.~(\ref{Frec2}). One can thus proceed to a 
numerical optimization with respect to the RS dependence, along the
same line as in section 2.5. The only difference is the Pad\'e
approximant form to be used: 
from eq.~(\ref{Fpiansatz}), for $m^{'''} \to 0$ the simple
pole behavior is now given by the {\em second} perturbative term, 
with coefficient $\alpha^{'}_\pi$, while the first perturbative 
term, $1$,
gives vanishing contribution for $m^{'''} \to 0$. A Pad\'e approximant
taking into account these properties, and using the full  
information as contained in $\alpha^{'}_\pi$, $\beta^{'}_\pi$ is 
\beq
(1 +{\alpha^{'}_\pi(a)\over F^{'}}+{\beta^{'}_\pi(a)
\over{(F^{'})^2}}
\;+\cdots) \; \simeq \frac{1+\lambda/(F^{'})^2}{1+\rho/F^{'}}\;
\equiv P_{2,1}(F^{'})\;,
\label{PadeFpi}
\eeq
which gives the result
\beq
F^2_{\pi,Pad\acute{e}}(\cdots) 
\raisebox{-0.4cm}{~\shortstack{ $=$ \\ $ m'' \to 0 $}}
\Lam\;
(2C)^{-2C} \;a^2\; exp\{\frac{2A_1}{2b_0}\}\;2b_0\:\delta_{\pi }
 \left[ 
\alpha^{'}_{\pi}(a,\cdots) -\frac{\beta^{'}_{\pi}(a,\cdots)}{
\alpha^{'}_{\pi}(a,\cdots)} \right]
\label{Fpipade}
\eeq
where the dots denote the RS dependence. 
The region around the optimum of (\ref{Fpipade}) 
with respect to the RS parameters is
illustrated  
in Fig.~2. In that case again we found a non-trivial flattest extrema
minimizing the curvature (plateau), corresponding 
e.g. for the $n_f=2$ case
to $a_{opt}\simeq 4.21$, $B_{1,opt}\simeq -0.017$.
The optimum
values are  
\beq
F^{Pad\acute{e}}_\pi(opt) \simeq 0.55\;\Lam(2)\;\;\;
(0.59\;\Lam(3)\;)\;,
\label{Fpinum}
\eeq
for $n_f =$ 2 (3) respectively.
This is also  
summarized in Table 1, 
section 5. From (\ref{Fpinum}), with $F_\pi \simeq $ 92 MeV (and
neglecting any explicit symmetry breaking effects due to the non-zero
$u, d, s$ masses), one thus obtains
$\Lam(2) \simeq 167$ MeV and $\Lam(3) \simeq 156$ MeV. Note that these
relatively low values are more in agreement with earlier estimates
of $\Lam$~\cite{sumrules,LamQCD} 
than with the recent experimental measurements 
of $\alpha_S(M_Z)$ at LEP~\cite{alphasexp}, 
the latter giving a larger $\Lam(3)$ if evolving
from the $M_Z$ scale down  
to very low $Q^2$  with naive perturbation theory. This does
not in principle exclude our results, due to the still
large uncertainties on $\Lam(2)$ and $\Lam(3)$ at very low
energies. Moreover, as pointed out in ref.~\cite{Shifman}, 
it could
always be that new physics contribution, e.g. supersymmetry, affects
the running of $\alpha_S$ between the $M_Z$ scale and low energy. \\
Finally, like in the mass case, we also give for comparison
the (a priori ambiguous) corresponding value of 
(\ref{Fpipade}) in the original $\MS$ scheme: 
\beq
F^{Pad\acute{e}}_\pi(\MS) \simeq 0.40\;\Lam(2)\;\;\;\;
(0.64\;\Lam(3)\;)\;.
\label{FpinumMS}
\eeq

\section{$\qq$}
\setcounter{equation}{0}
\subsection{Renormalization group invariant $m \qq$ ansatz}
The (bare) quark-antiquark condensate is most conveniently 
defined as
\beq
\qq  \equiv -\frac{i}{n_f} 
\raisebox{-0.4cm}{~\shortstack{ $lim $ \\ $ x \to 0 $}}
\;\;Tr S(x) 
\label{qqdef}
\eeq
where $S(x) = i<0\vert T \bar q(0) q(x) \vert 0 >$ is the quark
propagator~\footnote{
An equivalent definition is from the
derivative with respect to $m_0$ of the vacuum energy.
Note that in the $m \to 0$ limit, there is an arbitrary phase
in the definition (\ref{qqdef}):  
the sign of $\qq $ is thus fixed
a posteriori, by requiring that for real $ m > 0$,
$\qq \leq 0 $, to be consistent e.g. with the Gell-Mann Oakes
Renner~\cite{GOR} familiar relation. In what follows we shall designate
by $\qq $ only the {\em magnitude} of the quark condensate.}.
While the above definition (\ref{qqdef}) 
is gauge-invariant, as is well known 
$\qq $ is not separately RG invariant, 
but $m \qq $ is~\cite{Collins}, 
which is thus the appropriate 
quantity to consider when applying
the formalism developed in section 3.1.  
Once subtracting the divergences remaining after mass and coupling
renormalization, by 
following a construction similar to what was done 
for $F_\pi$, we obtain a finite ansatz 
for $\bar m \qq $. An awkward situation is that 
$\qq $ cannot be directly accessed, and has to be 
extracted from tiny explicit symmetry breaking effects due to $m \neq 0$. 
This is of course a well-known problem, not specific to our construction.
We shall discuss below in section 4.2 a possible way of extracting
a $\qq(\mu)$ value from our construction.\\ 
The perturbative expansion up to two-loop order for $\bar m \qq $
for $\bar m \neq 0$, calculated first in \cite{Spiridonov} 
and independently by us, is given explicitly in Appendix C.
Note that the three-loop order is not known, 
which is rather unfortunate since,
according to the derivation in section
3.1, it implies that one only knows unambiguously  
the first coefficient of $1/F$ in the $m\qq$ ansatz. 
To nevertheless give a more accurate estimate,  
following the usage in such a case, 
we take into account in our calculations  
the known RG dependence at 
${\cal O}(1/F^2)$~\footnote{ 
This allows at least some comparison with the true 
first order estimates, defined as the expressions obtained 
in the limit $b_1 = \gamma_1 = 0$.}.
Again,
 one assumes that the optimization with respect to the RS parameters
$a$, $\Delta\gamma_1$ 
etc (whose exact dependence at ${\cal O}(1/F^2)$ is also known) 
partly takes into account this ignorance on higher order
terms.\\ 

The resulting expression, including the full RS dependence, reads 
\bea
& 
{\ds
 {\bar m \qq \over{\Lam^4}} = \exp[\frac{4A_1}{2b_0}]\; 2b_0\;
{2^{-4 C} a^4\over{2 i \pi}} \oint {dy\over y}\; y^4 {e^{y/m^{'''}}}
\frac{1}{F^{'\;4 A^{'}-1} [C + F^{'}]^{\;4 B^{'}}} 
}
\; \times \nn \\
& 
{\ds
 {\delta_{\qq }
 \left(1 +{\alpha^{'}_{\qq}(a)\over{F^{'}}}+{\beta^{'}_{\qq}(a)
\over{F^{'2}}}
\;+\cdots \right)} 
}
\label{qqansatz}
\eea
with 
\beq
\delta_{\qq} = \frac{N_c}{4\pi^2}\;\frac{1}{2\gamma_0 -b_0}\; ,
\eeq
\beq
\alpha^{'}_{\qq}(a, \Delta\gamma_1) = 
\frac{1}{2b_0} 
\left[{4\over 3}(2\gamma_0 -b_0) -\pi^2(2\gamma^{'}_1 -b_1)-2(b_0 -
2\gamma_0)\;\ln\:a \right]
\label{alphapsi}
\eeq
and
\bea
& \beta^{'}_{\qq}(a, \Delta\gamma_1, B_1) = \frac{1}{(2b_0)^2}\left[
f^{(2)}_{\qq}(\Delta\gamma_1) +4\gamma_0(
2\gamma_0-b_0)
\ln^2\:a -10\frac{(2\gamma_0-b_0)}{6 \pi^2} \ln\:a \right] 
\nn \\ & \mbox{} -\frac{4 B_1}{2b_0} \alpha_{\qq}(a, \Delta\gamma_1)
 +\frac{B_1}{(2b_0)^2}[\frac{10}{3}(2\gamma_0-b_0)-4
\pi^2(2\gamma_1 -b_1)]\; ,
\label{betapsi}
\eea
where 
$f^{(2)}_{\qq}$ is given in Appendix C. 
\subsection{Explicit chiral symmetry-breaking corrections}
Clearly what we are really interested in is the value of the condensate 
in the exact chiral limit, $\bar m \to 0$.
Closely related to the latter problem, it is desirable to obtain a
connection between  
our construction and the familiar Gell-Mann--Oakes 
Renner (GOR) relation~\cite{GOR}, e.g. in the SU(2) case, 
\beq
-(m_u +m_d) \langle \bar u u\rangle \simeq F^2_\pi m^2_\pi 
+{\cal O}(m^2_{u,d}),
\label{GOR}
\eeq
where $\langle \bar u u\rangle = \langle \bar d d\rangle$ 
for exact SU(2) isospin symmetry, 
$F_\pi \simeq$ 92 MeV
and $m_\pi$ are the pion decay constant 
and mass respectively. \\ 
By definition 
in (\ref{GOR}) $m_{u,d}$ are the {\it current} masses, 
which break the chiral $SU(2)_L \times SU(2)_R$ symmetry explicitly. 
In contrast, as amply discussed, in (\ref{qqansatz}) 
$m^{'''}$ is an {\em arbitrary} parameter, destined to reach the 
chiral limit $m^{'''} \to 0$. 
Accordingly, $\bar m \to 0$
for $m^{'''} \to 0$,
so that one presumably expects only to recover the result
$\bar m \qq \to 0$ for $m^{'''} \to 0$. This is
actually the case:  
although
(\ref{qqansatz}) may potentially give a non-trivial result in the
chiral limit, typically the simple
pole residue ($\simeq 2b_0(2C)^{-C} \; \delta_{\qq }
\; \alpha_{\qq}(a)$,
upon neglecting higher-order
corrections), once we require
extrema of
this expression with respect to RS changes
(using for the $m''' \to 0$ limit
a Pad\'e approximant similar to the one for $F_\pi$),
we do {\em not} find
non-zero extrema. 
Therefore, such a result is not conclusive regarding
the actual value of $\qq (\bar\mu)$,
although it may be considered
a consistency cross-check
of our formalism. \\

Now, one possible way
of extracting $\qq(\mu) $ from the ansatz (\ref{qqansatz}) 
is the following. 
Let us introduce a small explicit chiral symmetry breaking mass term, 
\beq
-m_{0,exp} \bar q_i q_i\; ,
\eeq
to the basic Lagrangian (\ref{xdef}). After
renormalization, and carefully 
following the different steps as indicated
in section 2 and Appendix A, one can show that it amounts simply to
the following substitution into the integrand of the different previous
ansatzs:
\beq
y \to y + m^{''}_{exp}\; ,
\label{substexp}
\eeq
{\em except} in the factor $(dy/y)\; e^{y/m''}$, 
which remains unaffected,
and where 
$m''_{exp}$ is related to $\bar m_{exp}$
in exactly the same way as $m''$ is related to $\bar m$, see eq.
(\ref{msec2def}). Substituting (\ref{substexp})
into (\ref{qqansatz}), one can now expand
the right-hand side of (\ref{qqansatz}), 
firstly for small $(y+ m''_{exp})$,
and take in the resulting expression the true chiral 
limit, $y \to 0$. By finally subtracting $\bar m \qq $ on both
sides it gives an expression which depends on $m''_{exp}$ {\em only},
 in the
limit $m'' \to 0$. More precisely one obtains after simple 
algebra
\bea
 \bar m_{exp} \qq \equiv  
\raisebox{-0.4cm}{~\shortstack{ $lim $ \\ $ m'' \to 0 $}}
\left\{\;(\bar m +\bar m_{exp}) \qq 
-\bar m \qq \right\} 
\nn \\ 
 = \Lam^4 \; (2C)^{-4C}\delta_{\qq }\; 2b_0\; C^{\frac{C-B}{A}} 
 \exp[\frac{4A_1}{2b_0}] \;(m''_{exp})^{1/A}\;+{\cal O}(
 m^{''}_{exp})^{2/A}\;,
\label{qqexp}
\eea
in terms of the quantities already defined after eq.(\ref{qqansatz}),
including the RS dependence~\footnote{The $(1/A) \neq 1$ 
power of $m^{''}_{exp} \propto \bar m/\Lam$ in the second
line of 
(\ref{qqexp}) 
 may seem awkward at first. It is 
consistent however 
with the known perturbative behaviour of $\qq(\mu)$: 
from (\ref{qqexp}),
$\qq (\mu) \simeq m^{2b_0/\gamma_0-1} \ln (\mu/\Lam)$
for $\mu \gg \Lam$, so that 
since $m \simeq [\ln (\mu/\Lam)]^{-\gamma_0/2b_0}$
one recovers $\qq(\mu) \simeq [\ln (\mu/\Lam)]^{\gamma_0/2b_0}$
asymptotically, i.e. the condensate runs 
as inversely as the mass, as it should.}. \\ 

Using the relation between 
$m''_{exp}$ and $\bar m_{exp}$ in (\ref{msec2def}),
one can now cancel $\bar m_{exp}$ on both
sides of (\ref{qqexp}) and extract a value of $\qq $. 
Note that the dependence upon $\bar 
m_{exp} \equiv m_{exp}(\bar\mu)$ does
not completely cancel out, since  
$A(\Delta\gamma_1) \neq 1$ in a general
RS. Indeed in (\ref{qqexp}) there is an additional implicit dependence
on the scheme, through the relation between $m''_{exp}$ and $\bar m_{exp}$ 
in (\ref{msec2def}): this is expected, otherwise since 
$m'' \propto \bar m/\Lam $ one would  
obtain $\qq \propto\; pure~ number \times \Lam^3$,
which we know cannot be the case, due to the inherent $\mu$-dependence
of $\qq $. 
Consequently, in contrast with the dynamical
mass and $F_\pi$, it would not make much sense to invoke the PMS
with respect to the scale dependence $a$. Therefore, we shall fix 
(\ref{qqexp}) to its $\MS$ scheme value (i.e.
$a=1$, $\Delta\gamma_1 = B_1 =0$), which indeed makes comparison 
with other works easier, since the quark
condensate is generally
given in the $\MS$ scheme (and typically at $
\bar \mu =$ 1 GeV)~\cite{sumrules,SRqq}.
Although we do not try thus to numerically optimize 
expression (\ref{qqexp}), one should remember that
the ansatz (\ref{qqansatz}) 
is assumed to be already optimal with respect
to the variational mass parameter, $m''$. 
Indeed, the only remaining contribution from (\ref{qqansatz})
in (\ref{qqexp})
at first order in $m''_{exp}$ 
is the simple pole, but with residu given by the RS {\em independent}
zeroth
order terms, as a consequence of the $F^{-1}$ 
factor in (\ref{qqansatz}).
Accordingly,
potentially non-trivial
extrema, which may have appeared
 from the RS dependence at the next orders
(as was the case for $M_q$ and $F_\pi$)
are washed out. Taking into account the
next order $m''_{exp}$ terms in (\ref{qqexp}) one may eventually
obtain 
non-trivial extrema, but those small corrections 
are 
not expected to drastically change the 
results given below.\\

A more serious trouble when fixing the RS to $\MS$ 
is that the $\MS$ scheme
value $\Delta\gamma_1 =0$
is such that the extra singularities
occur at $Re[y] >0$, therefore rendering the expansion
around the origin $y \simeq 0$ a priori completely ambiguous, 
as explained in section 2. However, in eq.
(\ref{qqexp}), $y$ and $m''_{exp}$ are by definition independent
parameters,
so it turns out that the extra singularities,
as given by the zeros of
$dy/dF$, are independent of $m''_{exp}$ (equivalently,
because the extra singularities are
due to the Jacobian of the $F \to y$ transformation, 
and the $dy\:e^{y/m''}$ term remains unaffected by $m_{exp}$).
We thus
{\em assume} that the corresponding ambiguities 
cancel out
in the difference in expression (\ref{qqexp}). Although we were not able
to prove this statement rigorously,
at least it is easily checked that
the $\bar m \qq$ part in (\ref{qqexp}), 
which contains the singularities, do cancel in the first
order expansion in $m''_{exp}$. Within this assumption, 
we thus obtain the $\MS$ result:
\beq
\frac{\qq^{1/3}_{\MS}}{\Lam} (\bar \mu) \simeq 
0.647\; [\frac{\hat{m}(\bar \mu)}{\Lam(2)}]^{0.073} \; ; \;\;\;\;\;
 0.614 \;[\frac{\hat{m}(\bar \mu)}{\Lam(3)}]^{0.018}  ,
\label{qqnum}
\eeq
where the $\mu$ dependence is only due to the
remaining dependence on $\hat{m}(\bar \mu) \equiv 1/2(\bar m_u
+\bar m_d)$ or $\hat{m}(\bar \mu) \equiv 1/3(\bar m_u
+\bar m_d +\bar m_s)$, for $n_f=2$ or $n_f=3$ respectively.
Accordingly to extract a $\qq(\mu)$ value,  
we obviously have to use as input the 
explicit quark (current) masses $\hat{m}(\bar\mu)$. 
Since the only available values of the light quark masses
are extracted from a QCD sum rule determination of $\langle \bar s s
\rangle$~\cite{SRqq}, and using the $\hat{m}_{u,d}/\hat{m}_s$ 
mass ratio~\cite{Leutwyler} 
(and moreover assuming the validity of the
GOR relation), it is to be considered more as a consistency
check than a truly independent prediction. Note however that 
(\ref{qqnum}) is only weakly sensitive to the explicit breaking masses
$\hat{m}$, due to the small power coefficients involved.\\ 
Taking the latest results for the running masses at 1 GeV, as collected in 
ref.~\cite{Leutwyler}~\footnote{In (\ref{mdata2}), (\ref{mdata3})
the values
with errors added linearly were simply obtained from the errors quoted
in~\cite{Leutwyler}.},
\beq
\hat m (2) \equiv \frac{1}{2}(\bar m_u +\bar m_d)
\simeq 7.2 \pm 2.3 \mbox{MeV}\;
\label{mdata2}
\eeq
and
\beq
\hat m (3) \equiv \frac{1}{3}(\bar m_u +\bar m_d +\bar m_s)
\simeq 63.1 \pm 9.1 \mbox{MeV}\; ,
\label{mdata3}
\eeq
for $n_f=2, 3$ respectively, 
this gives
\beq
\qq^{1/3}_{\MS} (1 GeV) \simeq (0.500-0.525)\; \Lam (2)\;;
\;\;(0.602-0.606)\; \Lam (3)\;. 
\label{qqnum2}
\eeq

Alternatively, 
observing that in fact in (\ref{qqnum}) the power of $\hat{m}/\Lam $
only depends on the RS parameter $\Delta\gamma_1$, via the quantity $A$
defined in (\ref{defABC}), there may be a way of extracting
$\qq(\mu)$ in the exact chiral limit, 
$\bar m \to 0$, by simply choosing a RS
such that $A=1$ {\em exactly}. 
This occurs for $\Delta\gamma_1 = 
31/(288\pi^4)$ and $\Delta\gamma_1 = 5/(192\pi^4)$ for
$n_f =$ 2 and 3 respectively, and gives  
\beq
\qq^{1/3}_{\hat{m}\to 0} (1 GeV) \simeq 0.52 \;\Lam (2)\;,
\;\;0.58 \;\Lam (3)\;.
\label{qqnum0}
\eeq
Note that the results (\ref{qqnum2}) and
(\ref{qqnum0})
are fairly consistent, given the errors on the quark
mass $\hat{m}$ estimates and the presumed intrinsic error of our method.
The not so good agreement for $n_f=3$ may be attributed to the
fact that we have neglected the mass of the strange quark in our
(chiral symmetric) estimate of $F_\pi/\Lam$ and $\qq^{1/3}/\Lam$ 
(which is probably not
a very good approximation, as indicated e.g. from the fact
that $F_K/F_\pi \simeq 1.23$, experimentally.) \\ 
We thus obtain,  
within our approach, 
rather small $\qq$ values. 
To make our results consistent with the sum rules 
calculations~\cite{sumrules}, 
we would need
typically $\Lam \simeq$ 400 MeV, which 
contradicts the results in
section 3 for $F_\pi/\Lam$. 
(The former value of $\Lam$ is however more consistent
with the 
experimental results~\cite{alphasexp} from LEP,
$\alpha_S(M_Z) = 0.117 \pm 0.005$).\\

There are several ways in which 
our results may be interpreted. First, 
it is not excluded 
that our way of performing the small
expansion in $m_{exp}$ in (\ref{qqexp}), although mathematically
well-defined, may simply not be the physically sensible way of proceeding. 
Our framework amply relies on the continuation to very 
small values of the arbitrary mass parameter $m''$, thus it 
is may perhaps be only capable of determining the zero order
of a chiral symmetry breaking expansion in a small mass.
This is rather unfortunate for the quark condensate, 
which always 
appears 
as suppressed by an explicit mass in any physical relation.
Besides, we anyhow consider the results in (\ref{qqnum}),
(\ref{qqnum0}) on a less firm basis than the ones for $M_q$ and $F_\pi$
in (\ref{Mnum}) and (\ref{Fpinum}), 
due to the lack of PMS optimization (and due to 
the possible remaining ambiguities
of (\ref{qqansatz}) in the $\MS$ scheme as 
well~\footnote{As previously 
we can also choose the RS where  
$Re[y_{cut}] =0$ in (\ref{qqansatz}) and (\ref{qqexp}),  
which gives results in rough
consistency with those in (\ref{qqnum2}), (\ref{qqnum0}). However, 
due to the lack of RS invariance of $\qq$, it is then more difficult
to compare with the quoted results from 
other methods~\cite{sumrules}.}).

\section{Summary and discussion}
Let us first summarize our main results. \\ 

$\bullet$ We have shown that 
the variational expansion in arbitrary mass $\bar m$, as developped
in the context of the GN model~\cite{gn2}, can be formally  
extended to the QCD case. In particular we have seen how to
reconcile the variational expansion with renormalization, and
obtained a non-trivial, finite result in the chiral limit for the mass
gap, by resumming the variational series expansion in $x$ 
with an appropriate
contour integral. \\

$\bullet$ 
Next, we have exhibited the link between the finite variational 
ansatz and 
the RG solutions in renormalized form, with the identification
of specific fixed point
boundary conditions. 
This allows to generalize
the construction in a most straightforward way to 
the next RG orders and to more complicated quantities as well, such
as the relevant order parameters of
CSB, $\qq $ and $F_\pi$. \\

$\bullet$ We have studied in some details how the arbitrary
RS dependence influences our different ansatzs and, in particular,
how the extra singularities appearing in the
QCD case can be moved away by appropriate RS change at the second 
perturbative order.
The resulting
expressions in a generalized RS have been numerically optimized, within
a certain approximation, due to the complexity of the complete
optimization problem. The optimal values obtained
for $M_q/\Lam$ and $F_\pi/\Lam$, summarized
in table 1, are quite encouraging. Actually one should mention 
that the (unoptimized) corresponding
values in the original $\MS$ scheme, as given in eqs.~(\ref{MnumMS}), 
(\ref{FpinumMS}), would equally well be in reasonable agreement with other
estimates in the literature: given the large present uncertainties on
the values of $\Lam(2), \Lam(3)$ it is difficult to assert which 
results are closest to the largely unknow experimental values. 
However, the $\MS$ values are {\em a priori} ambiguous
due to the extra singularities, as discussed in section 2.4; 
moreover 
according to the PMS principle adopted here, 
we consider on general grounds 
the optimized results as being on a firmer basis. \\

$\bullet$ Finally, we have also introduced explicit chiral symmetry breaking
mass terms in our framework, which plays an especially
important role in extracting an estimate of $\qq$. It should also
be relevant to a more accurate evaluation of $F_\pi/\Lam$ and
$M_q/\Lam$ in the $SU(3)$ case, where the explicit symmetry breaking
due to the strange quark mass is presumably not 
negligible. We plan to study those explicit mass effects
in a more quantitative way in a future work. \\

\begin{table}[htb] \centering
\begin{tabular}{||l||l|l|l||}  \hline
 {\em Optimized } & {\em zero order} & {\em 1st order}& {2d order}  \\ 
{\em quantity}& ({\em pure RG}) & (optimized)    &    (optimized) \\ 
 & $n_f=2$, $n_f=3$ &$n_f=2$, $n_f=3$  &$n_f=2$, $n_f=3$  \\ \hline
 & & &  \\ 
$\frac{M^{Pad\acute{e}}}{\Lam}
 $ & 1.0,~~~~ 1.0   & 1.948,~~  1.948 &  2.97,~~   2.85 \\
  &   & $a_{opt}=e^{2/3}, e^{2/3}$   & $a_{opt} \simeq 1.42,~2.84$ \\ 
  &   &    & $B^{opt}_1 \simeq .12,~.06$
      \\ \hline
 & & &  \\ 
$\frac{F^{Pad\acute{e}}_\pi}{\Lam}
 $  & 0.355,~~ 0.355  & 0.359,~~ 0.360  &  0.55,~~   0.59   \\
  &   & $a_{opt}=e^{-1/12}, e^{-1/12}$   & $a_{opt} \simeq 4.2,~4.77$ \\ 
  &   &    & $B^{opt}_1 \simeq -.017,~-.026$
      \\ \hline
 & & &  \\ 
$\frac{(\bar m \qq^{Pad\acute{e}}
)^{1/4}}{\Lam}
 $ & 0.564,~~ 0.564 & 0.669,~~ 0.670 & $\simeq$ 0 ~~~~$\simeq$ 0      \\
  &   &  $a_{opt} = e^{5/12}, e^{5/12}$  &       \\ \hline
 & & &  \\ 
$ \frac{\qq^{1/3}}{\Lam} (\MS , \tilde{m})$ & 
0.65 $(\frac{\tilde{m}}{\Lam })^{0.07}$, 0.61 
$(\frac{\tilde{m}}{\Lam} )^{0.02}$ & -  & - \\
  &   &    &       \\ \hline
 & & &  \\ 
$ \frac{\qq^{1/3}}{\Lam} (\tilde{m}\to 0)$ & 
0.52,~~   0.58 & -  & -  \\
  &   &    &       \\ \hline 
\end{tabular}
\caption{Optimized results at different orders of the perturbative
(non-RG) corrections for $n_f=2$ and $n_f =3$ as function of the
basic QCD scale $\Lam$. The corresponding values of the RS parameters
at optima are indicated, when relevant. For $M/\Lam$ and
$F_\pi/\Lam$ the zero order corresponds to
neglecting completely the non-logarithmic corrections, 
and taking the simple pole resulting from (first order) RG dependence.
For $\qq^{1/3}/\Lam$ the ``zero" order corresponds in fact
to the (unoptimized) formula (\ref{qqexp}).} 
\label{table}
\end{table}
In table 1, we also give 
the results that are obtained when restricting the  
different ansatz to the first order, i.e. taking the limit
$b_1 =0$, $\gamma_1 =0$ and neglecting the $1/F^2$ terms in 
the different expressions for
$M^P_2$, $F_\pi$ and $m\qq$ respectively. (In that case, 
the only arbitrariness
is the renormalization scale, and the results were obtained 
from optimization with
respect to $a$). 
As well the ``zero order" results, i.e.  
obtained by taking only the {\em pure RG} dependence are shown.
Strictly speaking, those one-loop (and zero-loop) 
results are
again a priori ambiguous, being  
affected by the presence of extra singularities which cannot be moved away.
We nevertheless found useful to give them for completeness, 
as they were 
obtained from naively ignoring the ambiguities and taking the simple
pole residues in the relevant ansatzs. \\
 
Now another problem is how to estimate the error
of the method.
As may be expected,
the rigorous
theorems in ref.~\cite{JONES} cannot be applied to the present QCD
framework. Admittedly, our final numerical results thus require
a fair amount of mere ``trust", which is the usual 
problem with the PMS, even indeed when applied to 
perturbative expressions.
Note however that all our results are rigorously derived, as far 
as the pure RG behaviour is concerned: it is only the purely perturbative
non-RG corrections, ${\cal M}_{i} \neq 0$ in (\ref{contour8}), which
are treated according to the PMS. In other words we assume that
it is legitimate to treat the remaining perturbative
correction with a variant of the PMS.
In the GN model case~\cite{gn2},
the value of the curvature with respect to $a$
at the extremum gave a rather good qualitative
idea of the error, but there is a priori no more definite relationship
between the curvature at extrema points and the intrinsic error. 
The knowledge of the exact GN results allowed 
a comparison of different Pad\'e
approximant forms. As a partial qualitative cross-check,
we have tried a similar comparison in the present
case, and indeed found non-trivial optima 
(when those existed) of the same order of magnitude than
(\ref{Mnum}), (\ref{Fpinum}) 
with alternative Pad\'e constructions 
similar to those explained in 
ref.~\cite{gn2}~\footnote{In the GN model analysis,
we could not use a Pad\'e approximant like in (\ref{Mpade}),
which would give inconsistent results due to the fact that 
$b^{GN}_1 <0$.}. 
We do not find sensible
to make a systematic study in the QCD case
since, as above discussed, it would not
give much more quantitative idea on the intrinsic error of the method
itself.\\
 
Concerning the relatively low $\qq$ values, 
letting apart the possibility that
our method simply fails to extract an accurate 
estimate in the chiral limit,
for the reasons
discussed above, let us consider 
taking our $\qq$ results at face value and
discuss their consequences.
In fact, the 
possibility of a marginally small (or even zero) quark condensate was
recently raised in ref.~\cite{Sternetal,Stern}. 
Clearly, $\qq \neq 0$ is a sufficient but {\em not}
necessary condition for spontaneous CSB.  
For instance, as mentioned in section 3, $F_\pi$
is an equally well-defined order 
parameter. In ref.~\cite{Sternetal} 
the authors moreover emphasize
that there are at present no clear
experimental evidence for a realization of CSB 
through a large quark condensate. 
As stressed there, the familiar GOR 
relation is only the first order in $\bar m_{exp}$ expansion, assuming
the dominance of the $\qq$ condensate, precisely.
On the other hand, one may argue that 
there are some evidences, 
in particular 
from the spectral sum rules~\cite{sumrules} or more recently from 
lattice 
simulations~\cite{lattcsb}, 
of a low-energy QCD picture with a larger  
$\qq$. 
On the theoretical side, giving
up the dominance of the quark condensate 
gives a rather complicated framework
to describe low-energy QCD: although a generalization 
of chiral perturbation
theory is perfectly possible~\cite{Sternetal}, 
it looses a good part of its predictive power. Nevertheless it
is certainly not unreasonable to conclude
that a small quark condensate is at present not  
excluded from the  
data. \\
Another possibility would be simply that the dominant 
contribution to quark
 condensation has a very different origin than the 
mechanism leading to $F_\pi \neq 0$ and dynamical quark masses
$M_q \gg m_q$. 
As mentioned in the introduction other non-perturbative
effects associated with instantons, which  
are totally ignored here, are likely
to play some role in the CSB dynamics. For instance, it was argued  
long ago~\cite{CaDaGro} 
in some simplified picture of the QCD vacuum,
that the instanton/anti-instanton
interaction can give rise
to effective (non-local) four-quark interactions of the 
NJL type~\cite{NJL}. \\
In any event, the present construction  
may be considered a definite estimate of the 
contributions to order parameters 
that are associated
to the non-instantonic vacuum. 
\vskip 1 cm

{\bf Acknowledgements} \\

I am indebted to Andr\'e Neveu who first triggered my interest
in the subject and for many useful discussions. 
I am grateful to Eduardo de Rafael for 
valuable criticisms and remarks at a earlier stage of this work, 
and also to Chris Arvanitis, Fred\'eric Geniet, 
Georges Grunberg, Heinrich Leutwyler, Gilbert Moultaka, 
Jan Stern and Christoph Wetterich for helpful discussions.   

\newpage
\centerline{\large \bf Figure Captions.}
 
\vspace{30pt}
 
\noindent
Figure 1: dynamical quark mass $M_q/\Lam$ contribution from expression
(\ref{Mpade}) with $n_f =2$ versus the scale parameter $a$, for
different values of $B_1$ near the optimal (plateau)
region. The exact plateau corresponds to $B_1 =  
0.12$. $\Delta\gamma_1$ is fixed to $0.00437$, such that
$Re[y_{cut}] \simeq 0$. \\
\vskip 2 cm  
\noindent
Figure 2: dynamical $F_\pi/\Lam$ contribution from expression
(\ref{Fpipade}) with $n_f =2$ versus the scale parameter $a$, for
different values of $B_1$ near the optimal (plateau)
region. The exact plateau corresponds to $B_1 =
-0.0173$. $\Delta\gamma_1$ is fixed to $0.00437$, such that 
$Re[y_{cut}] \simeq 0$. 
 
\newpage 
\appendix
\section{Renormalization group material}
\setcounter{equation}{0}
We set here some definitions and normalization conventions 
for the renormalization group results
used at different stages of our construction. 

The {\em homogeneous} RG operator (i.e giving zero
when applied to a RG-invariant quantity) is 
taken as 
\beq
{\mu {d\over{d\mu}}} \equiv {\mu {\partial \over{\partial \mu}} }+
{\beta(g){\partial \over{\partial g}}} - \gamma_m(g)~m {\partial \over
{\partial m}},
\label{RGE}
\eeq
where
the RG coefficients are given by 
\beq
\beta(g) \equiv \mu {dg\over{d \mu}} = -b_0~ g^3 -b_1~ g^5 -b_2~ g^7 -...
\eeq
 
\beq
\gamma_m(g) \equiv -{\mu \over m}{dm \over{d \mu}} =
\gamma_0~ g^2 +\gamma_1~ g^4 +\gamma_2~ g^6 +...
\eeq
in terms of the coupling constant $g$, in a minimal subtraction scheme. 
In QCD ($\alpha_S \equiv g^2/(4 \pi)$)
the coefficients $b_i$ and $\gamma_i$ are known up to
the three-loop order\cite{b2} and read explicitly\cite{b0,b1,b2} 
in the $\MS$-scheme: 
\bea
b_0 & = & {1\over {16 \pi^2}}(11 -{2 \over 3} n_f)\;, \nn \\
 b_1 & = &{1\over {(16 \pi^2)^2}}(102 -{38\over 3} n_f)\;,
\nn \\
b_2 &= &{1\over {(16 \pi^2)^3}}
({2857\over 2} -{5033\over 18} n_f +{325\over 54} n^2_f)
\eea
\bea
\gamma_0 & = & {1\over{2 \pi^2}}\;, \nn \\
\gamma_1 & = & {1\over {(16 \pi^2)^2}}({404\over 3} -{40\over 9} n_f)\;,
\nn \\ \gamma_2 & = & {1\over {(16 \pi^2)^3}}
({7494 \over 3} +({320\over 3} 
\zeta(3) -{4432\over 27}) n_f -{280\over 81} n^2_f)
\eea
where $n_f$ is the number of active quark flavors. 

As is well-known~\footnote{
For a review on RG properties, see  
the excellent textbook by J.~C. Collins~\cite{Collins}.} 
only the first two $b_i$'s and $\gamma_0$ are 
RS independent. For a general perturbative 
change of scheme (restricted to second order which is sufficient for our
purpose),
\bea 
 g^2 \to g^{'2} = g^2\;(1 +A_1 g^2 +A_2 g^4 +\cdots)\; , \nn \\
 m \to m^{'} = m \;(1 + B_1 g^2 +B_2 g^4 + \cdots) \; , 
\label{RSchange}
\eea
the modification of the 
beta and gamma coefficients in the new (primed) scheme reads:
\bea
b^{'}_0 & = & b_0\; , \nn \\ 
b^{'}_1 & = & b_1\; , \nn \\
 b^{'}_2 & = & b_2 -A_1 b_1 +(A_2-A^2_1)b_0\; ;
\label{RSbeta}
\eea
and
\bea 
\gamma^{'}_0 & = & \gamma_0\; , \nn \\
 \gamma^{'}_1 & = & \gamma_1 +2b_0 B_1 -\gamma_0 A_1\; 
\equiv \; \gamma_1 +\Delta\gamma_1\;, \nn \\
\gamma^{'}_2 & = & 
\gamma_2 +2b_1 B_1 +2b_0(2B_2-B^2_1) -2 A_1 \gamma^{'}_1 -
\gamma_0 A_2\; .
\label{RSgamma}
\eea
Accordingly one can set $b^{'}_2 =0$ and $\gamma^{'}_2 =0$
by choosing appropriately 
$A_2$ and $B_2$ in (\ref{RSbeta}), (\ref{RSgamma}), for {\em arbitrary}
$B_1$ and $A_1$. 
These expressions are used in section 2, 3 and 4 to construct 
generalized RS dependent ansatzs, at the second perturbative order.\\ 
 
Next we define the leading log (LL), next-to-leading log
(NLL) etc, formal series for the
quantities we are interested in. Let us start with the mass:
from purely dimensional considerations, we write for the
RG invariant physical (pole)
mass,
\beq
M^P  = \bar m {\sum^\infty_{p=0} 
a_p({\bar m \over {\bar \mu}})~ \bar g^{2p}}
\label{MPRGsum}
\eeq
where
\beq
{a_p({\bar m \over {\bar \mu}}) \equiv {\sum^p_{r=0}
a_{p,r} [\ln ({\bar m \over \bar \mu})]^{p-r}}}\;,
\eeq
$\bar \mu $
is the $\MS$ scale and $\bar m \equiv m(\bar\mu)$
the
Lagrangian mass.
The $a_{p,0}$, ($p \geq 1$) 
coefficients of $[ln ({\bar m \over \bar \mu})]^p$
define the LL terms, the $a_{p,1}$ ($p \geq 2$)
coefficients of $[\ln ({\bar m \over \bar \mu})]^{p-1}$
define the NLL terms, etc. Accordingly the $a_{pp}$ 
coefficients are the
non-logarithmic perturbative terms, at order $p$. 
Considering now the
equation obtained by applying (\ref{RGE}) on
(\ref{MPRGsum}),
as giving the $\mu$-dependence of $M^P$
for {\it fixed} $\bar g^2$,
one obtains formal
series for the LL, NLL, etc~\cite{Collins}: \\ 
--the LL serie:
\beq
-p~ a_{p,0} = (\gamma_0 +2b_0(p-1)) a_{p-1,0}~~~~~~(p \geq 1;~~a_{0,0}
\equiv 1)
\label{LL}
\eeq
--the NLL serie:
\bea
(1-p)~ a_{p,1} &= (\gamma_0 +2b_0(p-1)) a_{p-1,1} +(\gamma_1 +2b_1(p-2))
a_{p-2,0} +\gamma_0 (p-1) a_{p-1,0},
\nn \\ &(p \geq 2)
\label{NLL}
\eea
and so on.
 
The universal (RS independent) LL series
can be easily resummed as
\beq
M^{LL}_1 = {m (1 +2b_0 g^2 L)^{-{\gamma_0 \over{2b_0}}} }.
\eeq
(where $L \equiv \ln ({\bar m \over \bar \mu})$).
In contrast the NLL, NNLL etc series are RS {\em dependent}, and not
straightforward to resum explicitly.
We have seen in section 2.3 how to find a resummation expression
for the NLL series
which can be checked to correctly
reproduce the terms formally defined in (\ref{NLL}) to all orders.
 
For of a composite operator ${\cal O}^n$
of naive mass
dimension n and depending only on $m$ and $g$,
as considered in section 3 and 4 for $F^2_\pi$ and $\bar m <\bar q
q>$ respectively~\footnote{We only consider 
however composite operators which can be
defined as two
point functions.}, it is straightforward to generalize the above 
expressions 
in (\ref{LL})--(\ref{NLL}). One thus obtains
similar $LL$ and $NLL$ series, that we do not display
explicitly. It is easily checked that the perturbative
expansion of the different
ansatzs do reproduce the correct formal series.

\section{A contour integral resumming the $x$ dependence}
\setcounter{equation}{0}
We review here the formalism originally introduced in
\cite{gn1,gn2} to obtain non-trivial result for the resummation
of the perturbative series in $x$, in the limit $x \to 1$. 
It leads to the dynamical 
mass ansatz in (\ref{contour77}), (\ref{contour7}) and its generalization 
for the composite operators in section 3, 4. \par
Consider the one-loop RG invariant (bare) expression for the 
mass $M_1$ as given in eq.~(\ref{Mbareansatz1}): 
\beq
M_1  = {m_0 \over{[1 -b_0 \Gamma(\epsilon /2)(4\pi)^{\epsilon/2} g^2_0
(M_1)^{-\epsilon} ]^{\gamma_0 \over{2b_0}} }}\; .
\label{Mbareansatz11}
\eeq
Performing the substitution 
\beq m_0 \to m_0 (1-x); ~~~~g^2_0 \to g^2_0 x,
\label{substitut2}
\eeq
provides a new quantity $M_1(x)$. 
To pick up the $x^q$ order term in $M_1(x) \equiv \sum^\infty _{q=0} 
a_q x^q $ (having in mind 
that we are actually interested in the limit $x \to 1$),
a convenient trick is by contour integration:
\beq
M^{(q)}_1 
\; \raisebox{-0.4cm}{~\shortstack{ $\to $ \\ $ (x \to 1)$}}
\; \sum^q _{k=0} a_k = 
{1\over{2 i \pi}} \oint dx (\frac{1}{x}+\cdots 
+\frac{1}{x^{q+1}})\;M_1(x)
.
\label{contour1} 
\eeq
Now performing the sum in (\ref{contour1})
 exhibits a $(1-x)^{-1}$ factor, 
cancelling the $(1-x)$ from (\ref{substitut2}). 
This results in the expression~\footnote{In (\ref{contour2})
there appeared in fact a factor of $1- x^{-(q+1)}$, from which only the last
term contributes to the integral due to the analyticity of
$f_0(x)$ defined in (\ref{f0def}).}:
\beq 
M^{(q)}_1 \;
\raisebox{-0.4cm}{~\shortstack{ $\to $ \\ $ (x \to 1)$}}
\; {1\over{2 i \pi}} \oint dx x^{-(q+1)}\;m_0 
[f_0(x)]^{-\frac{\gamma_0}{2b_0}}\;,
\label{contour2}
\eeq
where the contour is counterclockwise around the origin, and 
for convenience we defined the (recursive) function 
\beq
f_0(x) \equiv 1- b_0 \;x\; g^2_0 \Gamma[\frac{\epsilon}{2}] 
m^{-\epsilon}_0 (1-x)^{-\epsilon}\; (f_0)^{\epsilon \frac{\gamma_0}{2b_0}}\;,
\label{f0def} 
\eeq
dictated from eq.~(\ref{Mbareansatz11}).  
$f_0(x)$ has evidently a power series expansion in $x$, but,
less obviously, also admits an expansion in $(1-x)$, as noted by 
inverting its defining relation (\ref{f0def}). This implies in particular 
that 
$x =1$ is an (isolated) pole of $M_1$. \\ 
{\em Provided} that no extra singularities lie in the way, one may distort
the integration contour in (\ref{contour2})
to go around the cut lying along the
real positive axis and starting at $x =1$.
Actually,
 one can go a step further and reach the $q \to \infty$ limit:
after distorsion of the contour, only the vicinity of $x =1$ survives
for $q \to \infty$, that one can analyse by changing variable to
\beq
1 -x \equiv {v \over q}\;,
\eeq
and rescaling $m_0$ by introducing $m_0 = m^{'}_0 q$, 
keeping $m^{'}_0$ fixed as
$q$ goes to infinity. One finds in place of (\ref{contour2})
\beq
M_1\;
\raisebox{-0.4cm}{~\shortstack{ $=$ \\ $ (q \to \infty)$}}
\;{1\over{2 i \pi}} \oint {dv \over v} e^v
\; { v\; m^{'}_0 \over{f_0(v)^{\gamma_0
\over{2b_0}}} }
\label{contour3}
\eeq
where now $f_0(v) \equiv 1- b_0 \; g^2_0 \Gamma[\frac{\epsilon}{2}]
(m_0 v)^{-\epsilon} (f_0)^{\epsilon \frac{\gamma_0}{2b_0}}$. 
Once performing the renormalization
via $m_0 = \bar m Z_m$, $g^2_0 = \bar \mu^{\epsilon} Z_g 
g^2$,   $M_1$ 
in (\ref{contour3}) is now finite to all orders:
\beq
M_1 = {1\over{2 i \pi}} \oint dv e^{v} {\bar m \over {f^{{\gamma_0
\over{2b_0}}}}}
\label{contour5bis}
\eeq
where the
renormalized function 
\beq
f = 1 +2b_0 \bar g^2 
\ln[(\frac{\bar m v}{\bar \mu})
\;f^{-(\gamma_0/2b_0)}\;]. 
\eeq
We have thus shown how to recover finite quantities with a non-trivial
$x$ expansion. Eq.~(\ref{contour5bis}) 
however only includes the one-loop RG dependence.
To make the connection with the pole mass one should include 
the necessary non-logarithmic
perturbative corrections,  
already present e.g. at the one loop order in (\ref{mpertren}).
This can be done without affecting the contour integration
properties, except that the resulting expression of $M_1$ has a more
complicated structure around $v \simeq 0$, which can be 
however systematically
expanded around the origin in the way discussed in section 2.2.  
Generalization of the previous construction to 
the next RG order is possible~\cite{gn2} although here we shall
use in section 2.3 a more convenient construction directly in terms
of renormalized quantities.

\section{Perturbative results}
\setcounter{equation}{0}
In this section we collect for completeness the known perturbative expansions
for the three quantities of interest, $M(\bar m)$, $F^2_\pi(\bar m)$ and
$\bar m \qq (\bar m)$ respectively, in the $\MS$ scheme. These 
determine, among other things, 
the non-logarithmic perturbative corrections in our different ansatz,
as well as serve as a cross-check of the LL and NLL expansion properties.
\subsection{Quark mass}
  
The two-loop 
pole quark mass was calculated in ref.\cite{Gray}, with exact
dependence on the (current) 
quark masses running in the loops. It includes 
LL and NLL terms plus the non-logarithmic perturbative corrections.
As we consider only the case with $n_f$ equal mass
quarks, the results of \cite{Gray} take a simpler, entirely
analytical form:  
\beq
{M_Q\over{\bar m}} = {1 +{\gamma_0 ({2\over 3} -L) g^2}
 +[\gamma_0~ ({\gamma_0\over 2} +
b_0) L^2 +({\gamma^2_0\over 3} -{4\over 3} \gamma_0 b_0 -\gamma_1) L
+{K\over{(4 \pi^2)^2}} ] g^4}
\eeq
in terms of the (renormalized) current quark mass $\bar m$, 
with $L \equiv \ln(\bar m/\bar \mu)$, and 
\beq
K = {{\pi^2\over 9} \ln  2} +{7 \over 18} \pi^2 -{\zeta (3)\over{6}}
+{3673\over 288} -
({\pi^2\over 18} +{71\over 144}) n_f +{{(\pi^2 -3)}\over 8} (n_f -1)
-\frac{2}{3}\gamma^2_0 (4 \pi^2)^2
\eeq
for $n_f$ (equal mass) quarks
($\zeta (x)$ is the Riemann zeta-function). 
We refer to
\cite{Gray} for the details of this calculation.

\subsection{$F_\pi$ }
As discussed in section 3, the perturbative contributions to the pion decay
constant $F_\pi$ may be obtained from the axial-vector--axial-vector 
two-point
correlator
evaluated at $p^2  = 0$, where p is the external momentum. But that is
formally the
very same quantity which appears as the (neutral) part of the 
$\rho$-parameter (at $p^2 = 0$) 
in electroweak theory~\cite{Veltman}, up to some trivial overall factors. 
The two-loop pure QCD
corrections with exact quark mass dependence were first computed in
\cite{Abdel}, and recently the three-loop corrections in
\cite{Avdeev}. The only caution is to convert some of
these results, given in the on-shell scheme, to the
$\MS$-scheme\footnote{Results
in
$\MS$-scheme are also summarized 
in \cite{Avdeev}. Expression in eq.~(\ref{fpi2pert0})
 have been made consistent with
our definition of $\epsilon$, which differs from the one in  
\cite{Abdel,Avdeev}
by a factor of 2.}. More trivially,
those expressions 
were calculated for a top mass $m_t$, so that 
in the present context we 
replace $m_t \to  \bar m$, where $\bar m$
designates the 
current (equal) mass of the
light quarks. 
One thus has, for $D = 4 -\epsilon $
\bea
{F^2_{\pi,0}(pert)} = 
 -N_c {\bar m^2\over {2\pi^2} } \left\{ -{1\over \epsilon} +L \;\; +
{g^2\over{16 \pi^2}} \left[{8\over \epsilon^2} -{10\over {3 \epsilon}}
-8 L^2 -{4 \over 3} L -{1 \over 6} \right]
\right.  \nn \\ [0.3cm] 
\left.   \mbox{} +
({g^2\over{16 \pi^2}})^2 \left[ 
-{{{304\over 3} -{32 \over 9}n_f} \over \epsilon^3}
+{{132-{40 \over 9} n_f} \over \epsilon^2}
+{{-910/3 +24 \zeta(3) +16 n_f} \over { 9 \epsilon}} 
\right. \right. \nn \\
\left. \left. \mbox{}  +f^{30}_\pi   L^3 +
f^{31}_\pi L^2 +f^{32}_\pi L +f^{33}_\pi \right] \right\} 
\label{fpi2pert0}
\eea
where $L \equiv \ln (\bar m/\bar \mu)$ and
\beq
f^{30}_\pi = {304\over 3} -{32 \over 9}n_f
\eeq
\beq
f^{31}_\pi = -{136 \over 3} +{32 \over 9} n_f
\eeq
\beq
f^{32}_\pi = -8 \zeta(3)-{149\over 9}
 -{10 \over 9} n_f
\eeq
\bea
f^{33}_\pi = {16 \over 9}({51 \over 16} -36 \zeta(3) +27 \zeta(4) -6 B_4) +
         4 (3 +{28 \over 3} \zeta(3) -{27 \over 2} \zeta(4) +3 B_4) \nn \\
         + {4 \over 3} n_f (-{1\over 12} +{8\over 3} \zeta(3))
-{4 \over 3}(2 +12 \zeta(3))
\eea
with
\beq
B_4 = 16 Li_4({1\over 2}) +{2 \over 3} \ln ^4 2 -{2\over 3} \pi^2 \ln ^2 2
-{13 \over 180} \pi^4 = -1.7628..
\eeq
As explicit, (\ref{fpi2pert0}) 
still contains divergences even after mass and coupling renormalization,
and should be
renormalized by and additional subtraction, according to the 
discussion in section 3. 
Using eq.~(\ref{defsubtrac}) 
one then obtains after some algebra the perturbative 
coefficients of $1/F$ and $1/F^2$ 
respectively, $\alpha_{\pi}$ and $\beta_\pi $ given in 
(\ref{alphapi})--(\ref{betapi}), where the two-loop term $f^{(2)}_\pi$ 
is 
given as 
\bea
 f^{(2)}_\pi (\MS) =   \frac{1}{1152\pi^4(b_0 + \gamma_0)}
\left\{[455 -24 n_f +24\pi^2 
(\gamma_0+b_0)](\gamma_0-b_0) \right. \nn \\ 
\left. \mbox{}     +1152\pi^4(b_2-\gamma_2)
     + 960\pi^4(b_0-\gamma_0)\gamma_1
     +2304\pi^6 \gamma_1(\gamma_1-b_1)+ 
     36(b_0-\gamma_0)\zeta(3) \right\}
\eea  

\subsection{$\qq$}

We have computed independently 
the perturbative expression of the
quark-antiquark condensate $<q \overline{q}>$, with a non-zero mass, 
to two-loop order. 
Our results fully 
agree with the expressions first calculated by the authors of 
ref.~\cite{Spiridonov}.
With $D \equiv 4 -\epsilon$, one obtains in 
the $\MS$-scheme 
\beq
m_0 <\overline{q} q >^0_{pert} = {N_c {\bar m^4\over{2 \pi^2}}
 {[{1\over \epsilon} -L +{1\over 2}\;\; +
{g^2\over{\pi^2}} (-{1\over \epsilon^2} +{1\over{6 \epsilon}}
+L^2 -{5\over 6} L +{5\over 12} )]}}
\label{mqqpert0}
\eeq
where the index 0 stands for unrenormalized quantities. Like 
in the $F^2_\pi$ case, the expression in (\ref{mqqpert0}) has to be
further renormalized by a 
subtraction, which leads to the perturbative 
coefficients of $1/F$ and $1/F^2$ as given in (\ref{alphapsi}),
(\ref{betapsi}), where the 
two-loop $f^{(2)}_{<qq>}$ term~\footnote{We stress again that  
$f^{(2)}_{<qq>}$ is {\em not} the complete two-loop perturbative
term: the latter would need the knowledge of the 3-loop order $1/\eps$
coefficient, which we arbitrarily put to zero here.} is given as 
\bea
f^{(2)}_{<qq>}(\MS) = \frac{1}{6\pi^2(b_0 + 2\gamma_0)}\left[
-5 b_0^2 + 6\pi^2 (b_2 -
2\gamma_2) + 20\gamma_0^2 \right. \nn \\
\left. \mbox{}    +4\pi^2 b_0\gamma_1 - 12\pi^4 b_1\gamma_1 
- 4\gamma_1 +
24\pi^4\gamma_1^2\right]
\eea 

\newpage
\begin{figure}[htb]
\centerline{\epsffile{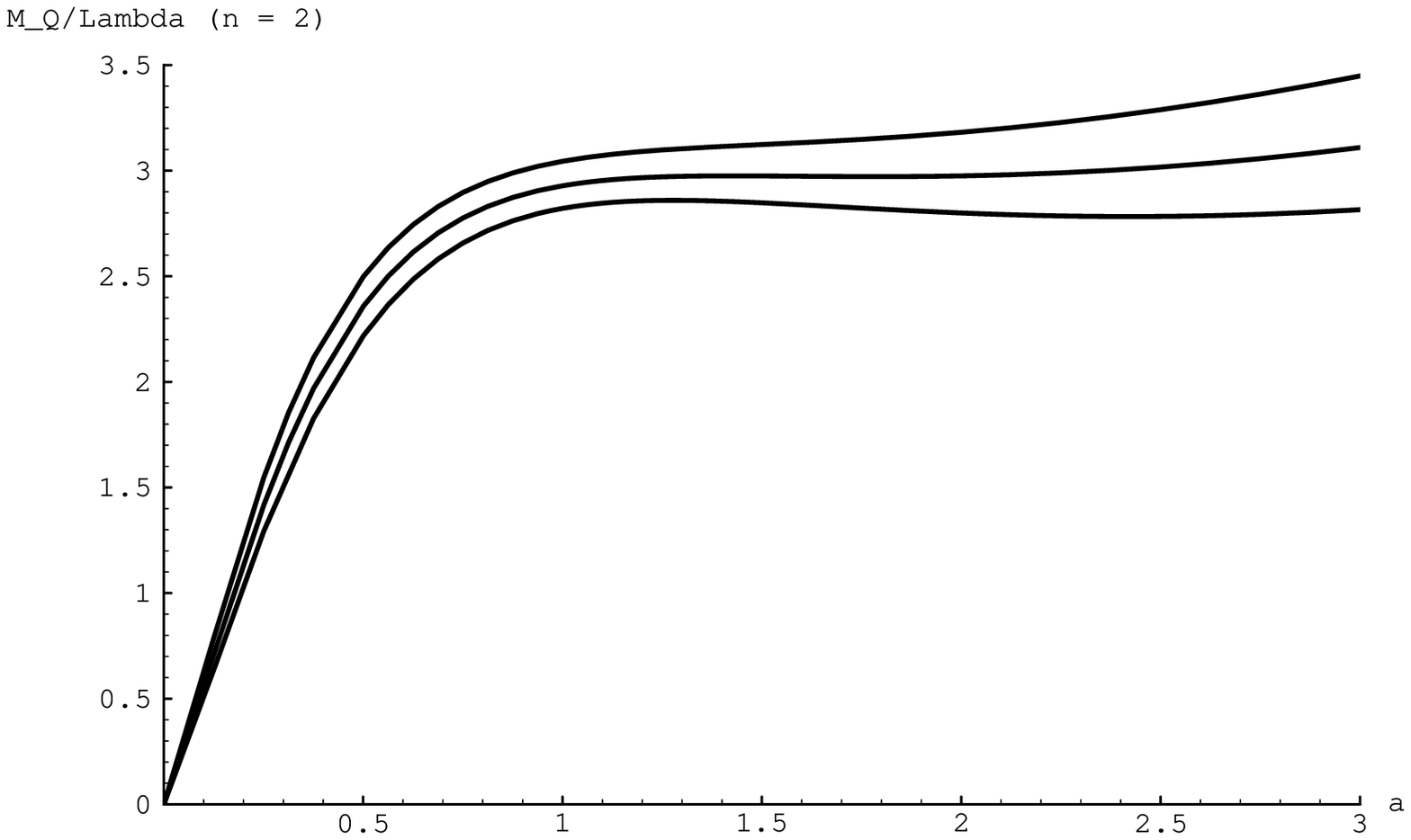}}
\vspace{-1.cm}
\caption{dynamical quark mass $M_q/\Lam$ contribution from expression
(\ref{Mpade}) with $n_f =2$ versus the scale parameter $a$, for
different values of $B_1$ near the optimal (plateau)
region. The exact plateau corresponds to $B_1 =
0.12$. $\Delta\gamma_1$ is fixed to $0.00437$, so that
$Re[y_{cut}] \simeq 0$.}
\label{fig1}
\end{figure}
 
\begin{figure}[htb]
\centerline{\epsffile{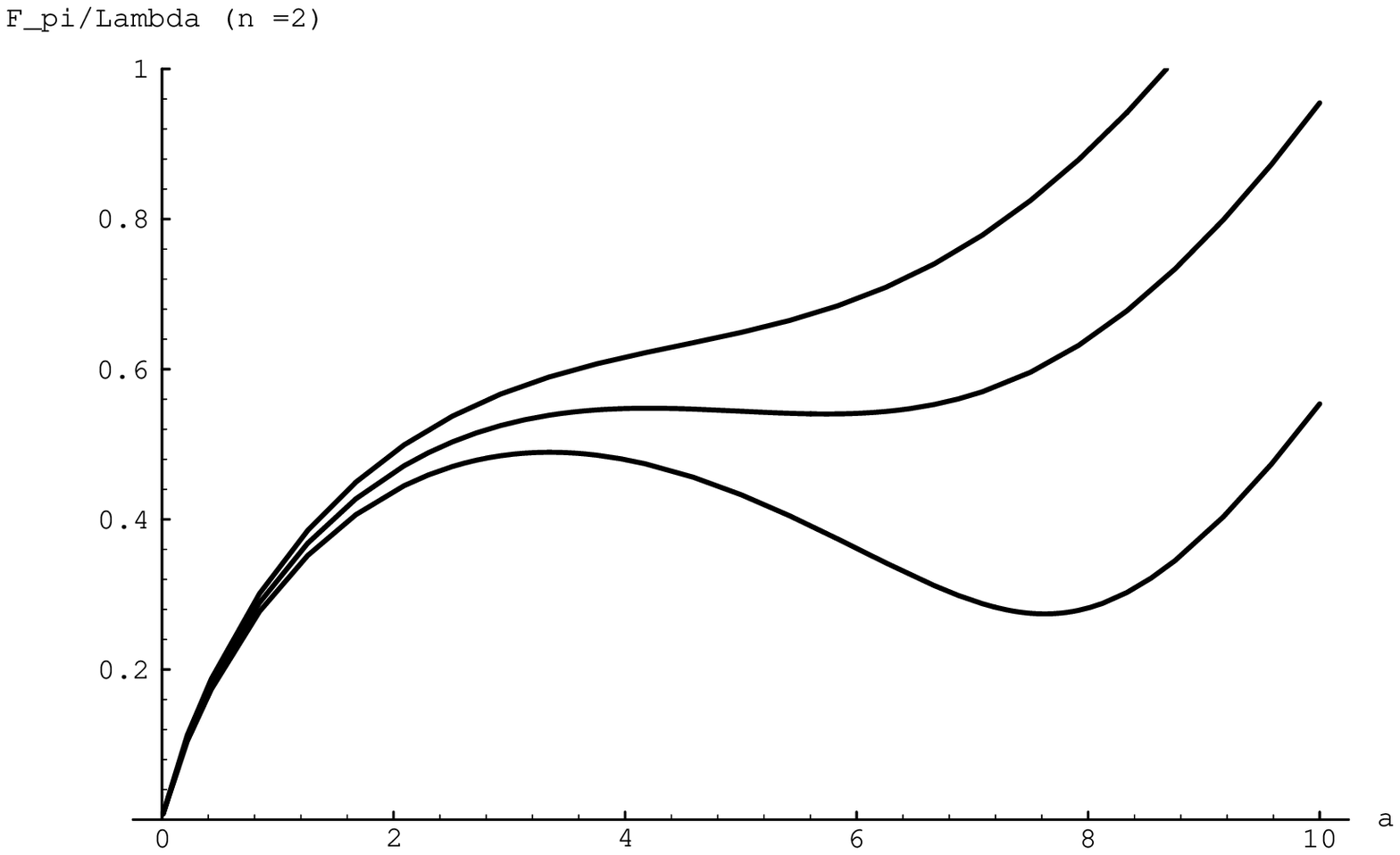}}
\vspace{-1.cm}
\caption{dynamical $F_\pi/\Lam$ contribution from expression
(\ref{Fpipade}) with $n_f =2$ versus the scale parameter $a$, for
different values of $B_1$ near the optimal (plateau)
region. The exact plateau corresponds to $B_1 =
-0.0173$. $\Delta\gamma_1$ is fixed to $0.00437$, so that
$Re[y_{cut}] \simeq 0$.}
\label{fig2}
\end{figure}

\end{document}